\documentclass[aps,prb,twocolumn,reprint,superscriptaddress,floatfix]{revtex4-2}

\usepackage{chemformula}
\usepackage{acronym} 
\usepackage{siunitx} 
\usepackage{braket}
\usepackage{hyperref}
\usepackage{orcidlink}

\DeclareSIUnit\angstrom{\text{\AA}}
\graphicspath{{./pictures/}}

\makeatletter
\patchcmd{\@outputpage@head}{\@ifx{\LS@rot\@undefined}{}{\LS@rot}}{}{}{}
\makeatother

\begin{document}

\title{X-ray Absorption Linear Dichroism at the Ti K-edge of Anatase \texorpdfstring{$\mathrm{TiO_2}$}{TiO2} Single Crystals}

\author{T.\ C.\ Rossi\orcidlink{0000-0002-7448-8948}}
\email{thomas.rossi@helmholtz-berlin.de}
\affiliation{Laboratory of Ultrafast Spectroscopy, Ecole Polytechnique F\'ed\'erale de Lausanne SB-ISIC, and Lausanne Centre for Ultrafast Science (LACUS), Station 6, Lausanne, CH-1015, Switzerland}
\author{D.\ Grolimund\orcidlink{0000-0001-9721-7940}}
\affiliation{Laboratory for Femtochemistry - MicroXAS beamline project, Paul Scherrer Institute, Villigen, CH-5232, Switzerland}
\author{ M.\ Nachtegaal\orcidlink{0000-0003-1895-9626}}
\affiliation{Bioenergy and Catalysis Laboratory, Paul Scherrer Institute, Villigen, CH-5232, Switzerland}
\author{O.\ Cannelli\orcidlink{0000-0002-1844-4799}}
\author{G.\ F.\ Mancini\orcidlink{0000-0002-7752-2822}}
\author{C.\ Bacellar\orcidlink{0000-0003-2166-241X}}
\author{D.\ Kinschel\orcidlink{0000-0002-0269-8567}}
\author{J.\ R.\ Rouxel\orcidlink{0000-0003-3438-6370}}
\affiliation{Laboratory of Ultrafast Spectroscopy, Ecole Polytechnique F\'ed\'erale de Lausanne SB-ISIC, and Lausanne Centre for Ultrafast Science (LACUS), Station 6, Lausanne, CH-1015, Switzerland}
\author{N.\ Ohannessian\orcidlink{0000-0002-3071-3373}}
\affiliation{Laboratory for Multiscale Materials Experiments, Paul Scherrer Institute, Villigen, CH-5232, Switzerland}
\author{D.\ Pergolesi\orcidlink{0000-0002-6231-0237}}
\affiliation{Laboratory for Multiscale Materials Experiments, Paul Scherrer Institute, Villigen, CH-5232, Switzerland}
\affiliation{Electrochemistry Laboratory, Paul Scherrer Institute, Villigen, CH-5232, Switzerland}
\author{T.\ Lippert\orcidlink{0000-0001-8559-1900}}
\affiliation{Laboratory for Multiscale Materials Experiments, Paul Scherrer Institute, Villigen, CH-5232, Switzerland}
\affiliation{Department of Chemistry and Applied Biosciences, Laboratory of Inorganic Chemistry, ETH Z\"urich, Vladimir-Prelog-Weg 1-5/10, 8093 Z\"urich, Switzerland}
\author{M.\ Chergui\orcidlink{0000-0002-4856-226X}}
\email{majed.chergui@epfl.ch}
\affiliation{Laboratory of Ultrafast Spectroscopy, Ecole Polytechnique F\'ed\'erale de Lausanne SB-ISIC, and Lausanne Centre for Ultrafast Science (LACUS), Station 6, Lausanne, CH-1015, Switzerland}

\date{\today}

\begin{abstract}
Anatase \ch{TiO2} (a-\ch{TiO2}) exhibits a strong X-ray absorption linear dichroism with the X-ray incidence angle in the pre-edge, the XANES and the EXAFS at the titanium K-edge. In the pre-edge region, the behaviour of the A1-A3 and B peaks originating from the 1s-3d transitions, is due to the strong $p$-orbital polarization and strong $p-d$ orbital mixing. An unambiguous assignment of the pre-edge peak transitions is made in the monoelectronic approximation with the support of \emph{ab initio} finite difference method calculations and spherical tensor analysis in quantitative agreement with the experiment. Our results suggest that several previous studies relying on octahedral crystal field splitting assignments are not accurate due to the significant \emph{p-d} orbital hybridization induced by the broken inversion symmetry in a-\ch{TiO2}. It is found that A1 is mostly an on-site 3d-4p hybridized transition, while peaks A3 and B are non-local transitions, with A3 being mostly dipolar and influenced by the 3d-4p intersite hybridization, while B is due to interactions at longer range. Peak A2 which was previously assigned to a transition involving pentacoordinated titanium atoms is shown for the first time to exhibit a quadrupolar angular evolution with incidence angle which implies that its origin is primarily related to a transition to bulk energy levels of a-\ch{TiO2} and not to defects, in agreement with theoretical predictions (Vorwerk \emph{et al} , Phys.\ Rev.\ B, \textbf{95}, 155121 (2017)). Finally, \emph{ab initio} calculations show that the occurence of an enhanced absorption at peak A2 in defect rich a-\ch{TiO2} materials originates from defect related $p$ density of states due to the formation of doubly ionized oxygen vacancies. The formation of peak A2 at almost the same energy for single crystals and nanomaterials is a coincidence while the origin is different. These results pave the way to the use of the pre-edge peaks at the \ch{Ti} K-edge of a-\ch{TiO2} to characterize the electronic structure of related materials and in the field of ultrafast X-ray absorption spectroscopy where the linear dichroism can be used to compare the photophysics along different axes.
\end{abstract}

\maketitle

\cleardoublepage

  \acrodef{XAS}{X-ray absorption spectroscopy}
  \acrodef{EXAFS}{extended X-ray absorption fine structure}
  \acrodef{XANES}{X-ray absorption near edge structure}
  \acrodef{MS}{multiple scattering}
  \acrodef{MO}{molecular orbital}
  \acrodef{FMS}{Full multiple scattering}
  \acrodef{FDM}{finite difference method}
  \acrodef{FDMNES}{finite difference method for near-edge structure}
  \acrodef{PLD}{pulse laser deposition}
  \acrodef{XRD}{X-ray diffraction}
  \acrodef{XRR}{X-ray reflectometry}
  \acrodef{DOS}{density of states}
  \acrodef{AF}{antiferromagnetic}
  \acrodef{LD}{linear dichroism}
  \acrodef{SI}{Supplementary Information}
  \acrodef{CBM}{conduction band minimum}

\section{Introduction}

Titanium dioxide (\ch{TiO2}) is one of the most studied large-gap semiconductor due to its present and potential applications in photovoltaics \cite{Freitag:2017di} and photocatalysis \cite{Nakata:2012hu}. The increasingly strict requirements of modern devices call for sensitive material characterization techniques which can provide local insights at the atomic level \cite{Suenaga:2000ix,Sherson:2010hg}. K-edge \ac{XAS} is an element specific technique, that is used to extract the local geometry around an atom absorbing the X-radiation, as well as about its electronic structure \cite{Milne:2014en}. A typical K-edge absorption spectrum usually consists of three parts: (i) in the high energy region above the absorption edge (typically $>\SI{50}{\electronvolt}$), the \ac{EXAFS}, contains information about bond distances. Modelling of the \ac{EXAFS} is rather straightforward, as the theory is well established \cite{Milne:2014en}; (ii) The edge region and slightly above it ($<\SI{50}{\electronvolt}$) represents the \ac{XANES}, which contains information about bond distances and bond angles around the absorbing atom, as well as about its oxidation state. In contrast to \ac{EXAFS}, \ac{XANES} features require more complex theoretical developments due to the multiple scattering events and their interplay with bound-bound atomic transitions; (iii) The pre-edge region consists of bound-bound transitions of the absorbing atom. In the case of transition metals, the final states are partially made of $d$-orbitals. Pre-edge transitions thus deliver information about orbital occupancies and about the local geometry because the dipole-forbidden $s$-$d$ transitions are relaxed by lowering of the local symmetry. The \ch{Ti} K-edge absorption spectrum of anatase \ch{TiO2} (a-\ch{TiO2}) exhibits four pre-edge features labelled A1, A2, A3 and B, while rutile \ch{TiO2} only shows three \cite{Brouder:2010go,Luca:2009bj}. Their assignment has been at the centre of a long debate, which is still going on, especially in the case of the a-\ch{TiO2} polymorph \cite{Brydson:1999bx,Uozumi:1992df,Wu:1997dm}. In this article, we use \ac{XAS} linear dichroism at the Ti K-edge to assign the pre-edge transitions of a-\ch{TiO2} since this technique can provide the orbital content in the final state of the bound transitions with the support of \emph{ab initio} \ac{FDM} calculations and spherical tensor analysis of the absorption cross-section.

Early theoretical developments to explain the origin of pre-edge features in a-\ch{TiO2} were based on \ac{MO} theory \cite{Fischer:1972gz,RuizLopez:1991cy,Grunes:1983gg} which showed that the first two empty states in a-\ch{TiO2} are made of antibonding $t_{2g}$ and $e_g$ orbitals derived from the $3d$ atomic orbitals of \ch{Ti}. Transitions to these levels have, respectively, been assigned to the A3 and B peaks while the absorption edge is made of \ch{Ti} $t_{1u}$ antibonding orbitals derived from \ch{Ti} $4p$ atomic orbitals. Although \ac{MO} theory can predict the energy position of the transitions accurately, it cannot compute the corresponding cross-sections and does not account for the core-hole to which quadrupolar transitions to $d$-orbitals at the K-edge are extremely sensitive \cite{Uozumi:1992df}. The corresponding transitions are usually red shifted by the core-hole and appear as weak peaks on the low energy side of the pre-edge. In a-\ch{TiO2}, peak A1 contains a significant quadrupolar component \cite{Uozumi:1992df}, sensitive to the core hole, which explains the inaccuracy of \ac{MO} theory to predict this transition. \ac{FMS} is a suitable technique to treat large ensembles of atoms and obtain accurate cross-sections \cite{RuizLopez:1991cy,Farges:1997kl,Wu:1997dm,Brydson:1999bx}. From \ac{FMS} calculations, a consensus has emerged assigning a partial quadrupolar character to A1, a mixture of dipolar and quadrupolar character with $t_{2g}$ orbitals to A3 and a purely dipolar transition involving $e_g$ orbitals to B \cite{RuizLopez:1991cy,Triana:2016fi}. However, as correctly pointed out by Ruiz-Lopez \cite{RuizLopez:1991cy}, this simple picture of octahedral symmetry energy split t$_{2g}$ and $e_g$ levels becomes more complicated in a-\ch{TiO2} because of the local distorted octahedral environment (D$_{2d}$ symmetry) which allows local $p-d$ orbital hybridization \cite{Yamamoto:2008bi}. In that case, the dipolar contribution to the total cross-section becomes dominant for every transition in the pre-edge region \cite{Cabaret:2010fp}. In addition, the cluster size used for the \ac{FMS} calculations has a large influence on the A3 and B peak intensities showing that delocalized final states (off-site transitions) play a key role in the pre-edge absorption region \cite{RuizLopez:1991cy}. Finally, the local environment around \ch{Ti} atoms in a-\ch{TiO2} is strongly anisotropic and \ch{Ti}-\ch{O} bond distances separate in two groups of apical and equatorial oxygens which cannot be correctly described with spherical muffin-tin potentials as implemented in \ac{FMS}. This limitation is overcome with the development of full potential \ac{FDM} calculations such as \ac{FDMNES} \cite{Joly:1999iq,Joly:2001fu,Joly:2009ha}.

Empirical approaches have been used by Chen and co-workers \cite{Chen:1997by} and Luca and co-workers \cite{Luca:1998dn,Hanley:2002go,Luca:2009bj} to establish correlations between the \ch{Ti} K pre-edge transitions in a-\ch{TiO2} and sample morphologies, showing that bond length and static disorder contribute to the change in the pre-edge peak amplitudes \cite{Chen:1997by} and that the A2 peak is due to pentacoordinated \ch{Ti} atoms \cite{Luca:1998dn,Hanley:2002go,Luca:2009bj}. Farges and co-workers confirmed this assignment with the support of \ac{MS} calculations \cite{Farges:1997kl}. The recent works by Zhang et al.\ \cite{Zhang:2008gt} and Triana et al.\ \cite{Triana:2016fi} have shown the strong interplay between the intensity of pre-edge features and the coordination number and static disorder, in particular in the case of the A2 peak. However, the A2 peak is also present in the \ac{XAS} of single crystals which suggests that the underlying transition is intrinsic to defect free a-\ch{TiO2}. Clear evidence of the nature of this transition is lacking which is provided in this work.

The clear assignment of the pre-edge features of a-\ch{TiO2} is important in view of recent steady-state and ultrafast \ac{XAS} \cite{RittmannFrank:2014fu,Santomauro:2016bg,Obara:2017bq} and optical experiments \cite{Baldini:2017kv}. In the picosecond \ac{XAS} experiments on photoexcited a-\ch{TiO2} nanoparticles above the band gap, a strong enhancement of the A2 peak was observed, along with a red shift of the edge \cite{RittmannFrank:2014fu}. This was interpreted as trapping of the electrons transferred to the conduction band at undercoordinated \ch{Ti} centres that are abundant in the shell region of the nanoparticles, turning them from an oxidation state of +4 to +3 \cite{RittmannFrank:2014fu}. The trapping time was determined by  femtosecond \ac{XAS} to be ca.\ $\SI{200}{\femto\second}$, i.e.\ the electron is trapped immediately at or near the unit cell where it was created \cite{Santomauro:2016bg,Obara:2017bq}. Further to this, the trapping sites were identified as being due to oxygen vacancies (\ch{O_{vac}}) in the first shell of the reduced \ch{Ti} atom. These \ch{O_{vac}}'s are linked to two \ch{Ti} atoms in the equatorial plane and one \ch{Ti} atom in the apical position to which the biexponential kinetics (hundreds of ps and a few ns) at the \ch{Ti} K-edge transient was attributed \cite{Santomauro:2016bg,Budarz:2017iu}. However, this hypothesis awaits further experimental and theoretical confirmation. In this sense, the assignment of peak A2 which provides the most intense transient signal in the pre-edge of a-\ch{TiO2} is a prerequisite. 

In this article, we provide a detailed characterization of the steady-state \ac{XAS} spectrum by carrying out a \ac{LD} study of anatase \ch{TiO2} single crystals at the \ch{Ti} K-edge, accompanied by detailed theoretical modelling of the spectra. We fully identify the four pre-edge bands (A1-A3 and B) beyond the octahedral crystal field splitting approximation used in several previous studies \cite{Wu:2004bs,Cabaret:2010fp}. Their dipolar and quadrupolar character is analyzed in detail as well as their on-site vs inter-site nature. The novelty resides in the quantitative reproduction of the experimental \ac{LD} data with \ac{FDM} calculations, the observation of the quadrupolar nature of peak A2 in agreement with theoretical predictions \cite{Vorwerk:2017gs} and the corresponding assignment of peak A2 as originating from a quadrupolar transition in single crystals and from defect states in nanomaterials. This delivers a high degree of insight into the environment of \ch{Ti} atoms, which is promising for future ultrafast X-ray studies of the photoinduced structural changes in this material.


\section{Experimental setup}


\subsection{Linear dichroism}

\label{exp_LD}

The \ac{LD} measurements are performed at the microXAS beamline of the SLS in Villigen, Switzerland using a double \ch{Si}(311) crystal monochromator to optimize the energy resolution. Energy calibration is performed from the first derivative of the \ac{XAS} spectrum of a thin \ch{Ti} foil. We used a moderately focused rectangular-shaped X-ray beam of $\SI{20}{}\times\SI{200}{\micro\meter\squared}$ in horizontal and vertical dimension, respectively. The \ac{XAS} spectrum is obtained in total fluorescence yield with a Ketek Axas detector system with Vitus H30 SFF and ultra-low capacitance Cube-Asic preamplifier (Ketek Gmbh).

The sample consists of a (001)-oriented crystalline a-\ch{TiO2} thin film of \SI{30}{\nano\meter} thickness. Sample growth and characterization procedures are reported in the \ac{SI} \S1 \footnote{See Supplementary Material at \url{https://journals.aps.org/prb/supplemental/10.1103/PhysRevB.100.245207/supplementary.pdf} for details about the sample synthesis and characterizations, the fitting procedure, the evolution of the A2, A3 and B peaks amplitudes with $\theta$ and $\phi$, the integrated DOS along $(x,y)$ and $z$ axes, the XAS spectra per equivalent sites, the comparison between calculated spectra with space group and supercell and the crystal symmetrization of spherical tensors.} (see also references \cite{Brouder:2008jc,AlsNielsen:2011vn} therein). Figure 1 shows a schematics of the sample motion required for the experiment. The sample was placed in the center of rotation of a system of stages which allow for both sample in-plane rotation ($\phi$) and orthogonal out-of plane rotation ($\theta$). By convention, a set of Euler angles $(\theta,\phi,\psi)$ orients the electric field $\hat\epsilon$ and wavevector $\hat k$ with respect to the sample. $\theta$ measures the angle between $\hat \epsilon$ and the $[001]$ crystal direction ($\hat z$ axis of the sample frame) orthogonal to the surface. $\phi$ measures the angle between $\hat\epsilon$ and the sample rotation axis $\hat x$. In principle, a third angle $\psi$ is necessary to fix the position of the wavevector in the orthogonal plane to the electric field but here $\psi=\SI{0}{\degree}$. The $\theta$ angles reported in the experimental datasets are with a maximum systematic offset of $\pm\SI{0.2}{\degree}$ which comes from the precision setting up the $\theta=\SI{0}{\degree}$ reference from the sample half-clipping of the X-ray beam at grazing incidence. The precision of the rotation stage of $\pm\SI{0.01}{\degree}$ is negligible with respect to this angular offset.


\begin{figure}
\includegraphics[scale=8.5]{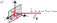}
\includegraphics[scale=8.5]{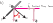}
\caption{Linear dichroism experiment with (a) side view and (b) top view. The sample surface is in grey while the incident X-ray beam is in pink. A set of Euler angles ($\theta,\phi,\psi$) is used to orient the electric field $\hat\epsilon$ and wavevector $\hat k$ of the incident X-ray beam with respect to the sample.}
\label{experimental_design}
\end{figure}


\ac{LD} is usually studied with the sample rotated in the plane orthogonal to the incident X-ray beam ($\phi$-rotation) \cite{Brouder:tj}. In this work, the novelty comes from the sample rotation around $\hat x$ ($\theta$-rotation) which provides the largest changes in the \ac{XAS}. This rotation induces a change of X-ray footprint onto the sample surface. We clearly show that it does not introduce spectral distortions because the effective penetration depth of the X-rays through the material (between 97 and \SI{580}{\nano\meter} across the absorption edge of a-\ch{TiO2} for the largest footprint at $\theta=\SI{1}{\degree}$ used here \cite{Henke:1993eda}) is kept constant as the sample is much thiner than the attenuation length at the \ch{Ti} K-edge. Instead, the total amount of material probed by the X-rays changes due to the larger X-ray footprint when $\theta$ increases and a renormalization over the detected number of X-ray fluorescence photons is required. This is done with the support of the \ac{FDMNES} calculations since a few energy points have $\theta$-independent cross-sections as previously reported on other systems \cite{George:1989ev,Loupias:1990de,Oyanagi:1987ik,Oyanagi:1989gz,Pettifer:1990kv,Stizza:1986ff,Fretigny:1986ea} (\emph{vide infra})\footnote{For a spectrum measured well above the absorption edge, the atomic background absorption converges for any incident polarization and can also be used in principle to renormalize the spectra.}. With this renormalization procedure performed at a single energy point (\SI{4988.5}{\electronvolt}), we could obtain a set of experimental points with $\theta$-independent cross-sections at the energies predicted by the theory confirming the reliability of the method. Hence, crystalline thin films with suitable thicknesses with respect to the X-ray penetration depth offer more possibilities to study \ac{LD} effects than single crystals and prevent the usual self-absorption distortion of bulk materials when using total fluorescence yield detection \cite{Carboni:2005jf}.

\begin{table*}[htp]
\tiny
\caption{Previous assignments of the final states involved in the pre-edge transitions of the \ch{Ti} K-edge \ac{XAS} spectrum of a-\ch{TiO2}. The orbitals with dominant contribution to the transition are emphasized in bold. $E1$ is for dipolar transitions and $E2$ for quadrupolar transitions. Off-site transitions are in red, on-site transitions are in black. Irreducible representations corresponding to the final state in the considered point group of a-\ch{TiO2} are in parentheses. The dash "-" symbol refers to orbital hybridization.}
\begin{center}
\begin{tabular}{|c|c|c|c|c|}
\hline
reference 				& A1 															& A2 														& A3 															& B \\
\hline
\cite{Wu:1997dm}		& $\boldsymbol{3d_{x^2-y^2}(b_1)}$, $4p_x,4p_y,3d_{xz},3d_{yz}(e)$				& $\boldsymbol{4p_z},\boldsymbol{3d_{xy}(b_2)}$, $4p_x,4p_y,3d_{xz},3d_{yz}(e)$	& $\boldsymbol{4p_z},\boldsymbol{3d_{xy}(b_2)}$, $3d_{z^2}(a_1)$					& $\boldsymbol{4p},\boldsymbol{4s}$ \\
\cite{Cabaret:2010fp}	& $E1$: $\boldsymbol{p(t_{2g})}$, $E2$: $3d(t_{2g})$							&															& $E1$: $\boldsymbol{p(e_g)}$, \textcolor{red}{$\boldsymbol{p-3d(t_{2g})}$}, $E2$: $3d(e_g)$			& \textcolor{red}{$E1$: $\boldsymbol{p_z},\boldsymbol{3d(e_g)}$} \\
\cite{Triana:2016fi} 		& $E1$: $\boldsymbol{4p-3d(t_{2g})}$, $E2$: $3d(t_{2g})$													& 															& $E1$: $\boldsymbol{4p}$,$\boldsymbol{d_{xy}}$,$\boldsymbol{d_{xz}}$,$\boldsymbol{d_{yz}}(\boldsymbol{t_{2g}})$ 			& $E1$: $\boldsymbol{4p}$,$\boldsymbol{d_{x^2-y^2}}$,$\boldsymbol{d_{z^2}}(\boldsymbol{e_g})$ \\
This work 				& $E1$: $\boldsymbol{4p_{x,y}}-\boldsymbol{3d_{xz}}$,$\boldsymbol{3d_{yz}}$, $E2$: $d_{xz}$,$d_{yz}$,$d_{x^2-y^2}$	& $E1$: $4p_z-3d_{xy}$, $E2$: $\boldsymbol{3d_{xy}}$, $3d_{z^2}$							& \textcolor{red}{$E1$: $\boldsymbol{4p_x}$,$\boldsymbol{4p_y}$,$\boldsymbol{4p_z}$--$3d_{xy},3d_{z^2}$}								& \textcolor{red}{$E1$: $\boldsymbol{4p_x}$,$\boldsymbol{4p_y}$,$\boldsymbol{4p_z}$} \\
\hline
\end{tabular}
\end{center}
\label{literature_assignment}
\end{table*}%


\section{Theory}

\subsection{Recent developments in computational methods}

Recently, there have been two main developments in the computation of \ac{XANES} spectra. The first is based on band structure calculations (LDA, LDA+U,\dots), which compute potentials self-consistently with and without the core-hole before the calculation of the \ac{XAS} absorption cross-section with a core-hole in the final state \cite{Cabaret:2010fp,Gougoussis:2009kr}. This approach provides excellent accuracy but is limited to the few tens of eV above the absorption edge due to the computational cost of increasing the basis set to include the \ac{EXAFS}. The second one, the \ac{FDMNES} approach implemented by Joly \cite{Joly:2009ha,Joly:2001fu}, overcomes the limitations of the muffin-tin approximation in order to get accurate descriptions of the pre-edge transitions especially for anisotropic materials. The recent theoretical work by Cabaret and coworkers combining GGA-PBE self-consistent calculations with \ac{FDMNES} \cite{Cabaret:2010fp} concluded that in a-\ch{TiO2}, peak A1 is due to a mixture of quadrupolar ($t_{2g}$) and dipolar transitions ($p-t_{2g}$), A3 to on-site dipolar ($p-e_g$), off-site dipolar ($p-t_{2g}$) and quadrupolar ($e_g$) transitions, while B is due to an off-site dipolar transition ($p_z-e_g$). These results, together with those of previous works are summarized in Table \ref{literature_assignment}. However, experimental support to the pre-edge assignments is still lacking, and is provided in this work using \ac{LD} \ac{XAS} at the \ch{Ti} K-edge of a-\ch{TiO2} with the theoretical support of \emph{ab-initio} full potential \ac{FDMNES} calculations and spherical harmonics analysis of the \ac{XAS} cross-section.

\subsection{Finite difference \emph{ab-initio} calculations}

The \emph{ab initio} calculations of the \ac{XAS} cross-section were performed with the full potential \ac{FDM} as implemented in the FDMNES package \cite{Joly:1999iq,Joly:2001fu}. A cluster of \SI{7.0}{\angstrom} was used for the calculation with the fundamental electronic configuration of the oxygen atom and an excited state configuration for the titanium atom (\ch{Ti}:\ [Ar]3d$^1$4s$^2$4p$^1$) as performed elsewhere \cite{Zhang:2008gt}. We checked the convergence of the calculation for increasing cluster sizes and found minor evolution for larger cluster radii than \SI{7.0}{\angstrom} (123 atoms). The Hedin-Lundqvist exchange-correlation potential is used \cite{Hedin:2001jx}. A minor adjustment of the screening properties of the $3d$ levels is needed to match the energy position of the pre-edge features with the experiment. We found the best agreement for a screening of 0.85 for the $3d$ electrons. After the convolution of the spectrum with an arctan function with maximum broadening of \SI{1.5}{\electronvolt}, a constant gaussian broadening of \SI{0.095}{\electronvolt} is applied to account for the experimental resolution of the experiment and get the closest agreement with the broadening of the pre-edge peaks.


\subsection{Spherical tensor analysis of the dipole and quadrupole cross-sections}

Analytical expressions of the dipole and quadrupole \ac{XAS} cross-sections ($\sigma^D(\hat\epsilon)$ and $\sigma^Q(\hat\epsilon,\hat k)$, respectively) are obtained from their expansion into spherical harmonic components \cite{Brouder:tj,Brouder:1990eo}. The expressions of $\sigma^D(\hat\epsilon)$ and $\sigma^Q(\hat\epsilon,\hat k)$ depend on the crystal point group which is D$_{4h}$ ($4/mmm$) for a-\ch{TiO2}. The dipole cross-section is given by:
\begin{equation}
\sigma^D(\hat\epsilon)=\sigma^D(0,0)-\frac{1}{\sqrt{2}}(3\cos^2\theta-1)\sigma^D(2,0)
\label{dipolar_dichroic}
\end{equation}
and the quadrupole cross-section by:
\begin{equation}
\begin{split}
\sigma^Q(\hat\epsilon,\hat k)=\sigma^Q(0,0)\\
+\sqrt{\frac{5}{14}}(3\sin^2\theta\sin^2\psi-1)\sigma^Q(2,0) \\
+\frac{1}{\sqrt{14}}[35\sin^2\theta\cos^2\theta\cos^2\psi\\
+5\sin^2\theta\sin^2\psi-4]\sigma^Q(4,0) \\
+\sqrt{5}\sin^2\theta[(\cos^2\theta\cos^2\psi\\
-\sin^2\psi)\cos4\phi-2\cos\theta\sin\psi\cos\psi\sin4\phi]\sigma^{Qr}(4,4)
\end{split}
\label{quadrupolar_dichroic}
\end{equation}
with $\theta$, $\phi$ and $\psi$ as defined in the \ch{Ti} site point group (D$_{2d}$). $\sigma^X(l,m)$ with $X=D,Q$ is the spherical tensor with rank $l$ and projection $m$. $\sigma^{Xr}$ refers to the real part of the cross-section. The Euler angles $(\theta,\phi,\psi)$ in the experiment are referenced to the crystal frame which is rotated in the $(O,\hat x,\hat y)$ plane with respect to the Euler angles in the \ch{Ti} site frame. Consequently, the angles in equations \ref{dipolar_dichroic} and \ref{quadrupolar_dichroic} differ from the angles defined in Figure \ref{experimental_design} by a rotation of $\phi$. In the \ch{Ti} site frame, the $\hat x$ and $\hat y$ axes are bisectors of the \ch{Ti-O} bonds while the crystal frame is along the bonds. The matrix $R$ to go from the site frame to the crystal frame is,
\begin{equation}
R=\begin{pmatrix}
\frac{1}{\sqrt{2}} & \frac{1}{\sqrt{2}} & 0 \\
-\frac{1}{\sqrt{2}} & \frac{1}{\sqrt{2}} & 0 \\
0 & 0 & 1 \\
\end{pmatrix}
\end{equation}
In the following, the polarizations of $\hat\epsilon$ and $\hat k$ are given in the crystal frame. Consequently, the corresponding polarizations for the site frame are given by $\hat\epsilon_S=R^{-1}(\hat \epsilon)$ and $\hat k_S=R^{-1}(\hat k)$.  

Although some terms of $\sigma^D(\hat\epsilon)$ and $\sigma^Q(\hat\epsilon,\hat k)$ may be negative, the total dipolar and quadrupolar cross-sections must be positive putting constraints on the values of $\sigma^D(l,m)$ and $\sigma^Q(l,m)$. The electric field $\hat\epsilon$ and wavevector $\hat k$ coordinates in the $(\hat x,\hat y, \hat z)$ basis of Figure \ref{experimental_design} are given by:
\begin{equation}
\hat\epsilon=\begin{pmatrix}
\sin\theta\cos\phi \\
\sin\theta\sin\phi \\
\cos\theta
\end{pmatrix},\text{ }
\hat k=\begin{pmatrix}
\cos\theta\cos\phi \\
\cos\theta\sin\phi \\
-\sin\theta
\end{pmatrix}.
\end{equation} 
Hence the detail of the cross-section angular dependence in equations \ref{dipolar_dichroic} and \ref{quadrupolar_dichroic} requires the estimate of the spherical tensors $\sigma^D(l,m)$ and $\sigma^Q(l,m)$ as performed elsewhere \cite{Brouder:2008jc}. The \ac{XAS} cross-section measured experimentally is an average over equivalent \ch{Ti} atoms under the symmetry operations of the crystal space group. The analytical formula representing this averaged cross-section requires the site symmetrization and crystal symmetrization of the spherical tensors, which is provided in \ac{SI} \S7 and \S8 \cite{Note1}. From this analysis, we obtain nearly equal (up to a sign difference) crystal-symmetrized ($\braket{\sigma(l,m)}_X$), site-symmetrized ($\braket{\sigma(l,m)}$) and standard ($\sigma(l,m)$) spherical tensors. Assuming pure $3d$ and $4p$ final states in the one-electron approximation, analytical expressions are provided for $\sigma^D(\hat\epsilon)$ and $\sigma^Q(\hat\epsilon,\hat k)$ whose angular dependence with $\theta$ and $\phi$ are given in Table \ref{dipole_table}. The full expressions of the cross-sections are provided in \ac{SI} \S7 \cite{Note1}. In this paper, we analyze the angular dependence of the pre-edge peak intensities with $\theta$ and $\phi$ and assign them to specific final states corresponding to \ch{Ti}-$3d$ and/or $4p$ orbitals with the support of both \ac{FDM} and spherical tensor analysis.

\begin{table*}[htp]
\caption{Angular dependence along $\theta$ and $\phi$ of the dipole and quadrupole \ac{XAS} cross-section dominant terms in the equations (\ref{dipolar_dichroic}) and (\ref{quadrupolar_dichroic}) at the \ch{Ti} K-edge of a-\ch{TiO2} according to the final state of the transition. Transitions to $p$ final states are dipole allowed ($\sigma^D(\hat\epsilon)$) while transitions to $d$ final states are quadrupole allowed ($\sigma^Q(\hat\epsilon,\hat k)$).}
\begin{center}
\begin{tabular}{|c|c|c|}
\hline
\textbf{final state} & $\boldsymbol{\sigma^D(\hat\epsilon)}$ or $\boldsymbol{\sigma^Q(\hat\epsilon,\hat k)}$ $\theta$-dependence & $\boldsymbol{\sigma^D(\hat\epsilon)}$ or $\boldsymbol{\sigma^Q(\hat\epsilon,\hat k)}$ $\phi$-dependence \\
\hline
$p_x,p_y$ 					& $-\cos^2\theta$ 	& no dependence		\\
$p_z$ 						& $\cos^2\theta$ 	& no dependence		\\
$d_{z^2}$	& $\sin^2\theta\cdot\cos^2\theta$ & no dependence	\\
$d_{xy}$	& $\sin^2\theta\cdot\cos^2\theta$ & $\cos(4\phi)$	\\
$d_{x^2-y^2}$	& $\sin^2\theta\cdot\cos^2\theta$ & $-\cos(4\phi)$	\\
$d_{xz}$,$d_{yz}$ 				& $-\sin^2\theta\cdot\cos^2\theta$  & no dependence	\\
\hline
\end{tabular}
\end{center}
\label{dipole_table}
\end{table*}%


\section{Results}

\label{results_section}

\begin{figure*}
\begin{center}
\includegraphics[scale=0.30,trim={0 0 0 0},clip]{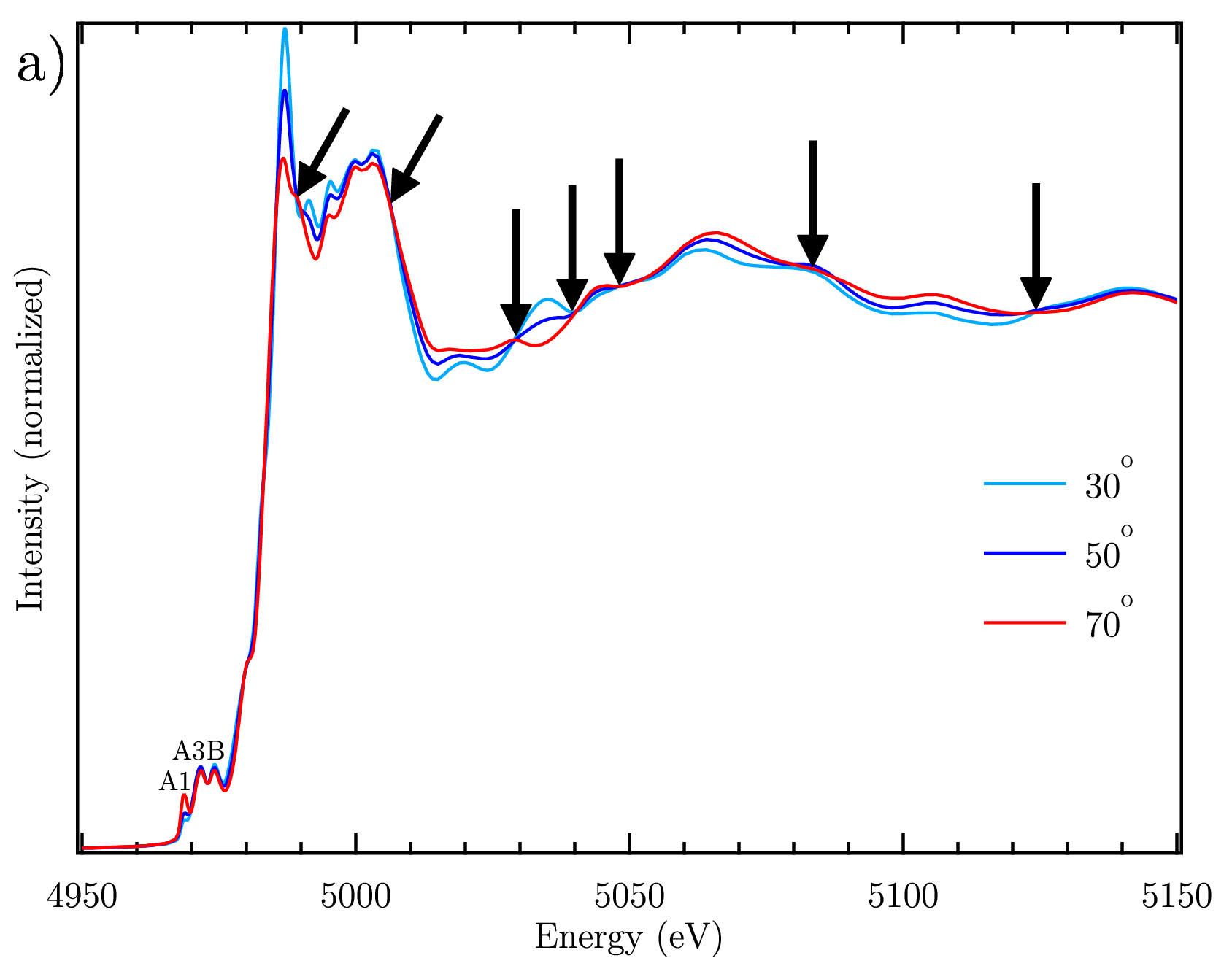}
\includegraphics[scale=0.30,trim={0 0 0 0},clip]{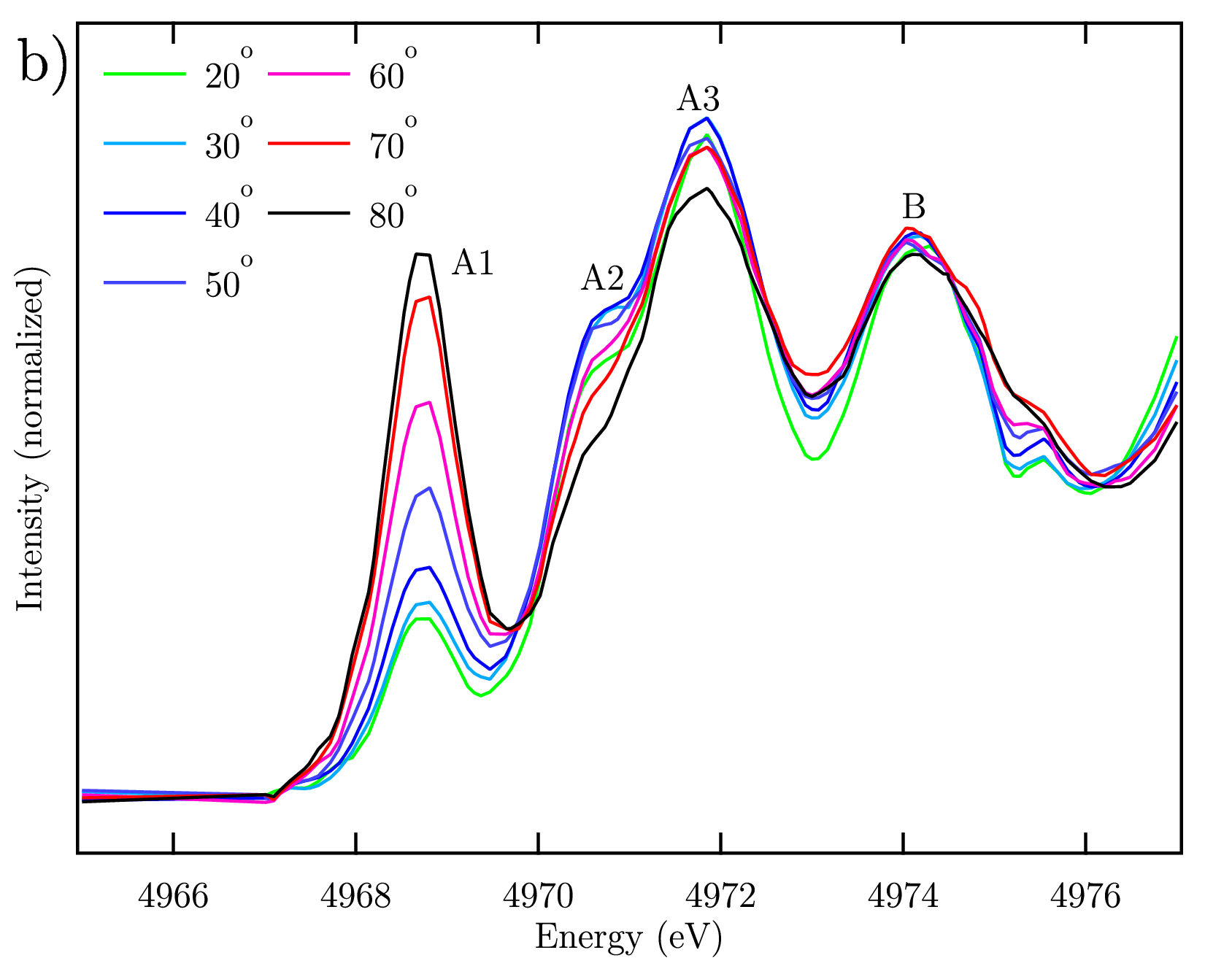}
\end{center}
\begin{center}
\includegraphics[scale=0.30,trim={0 0 0 0},clip]{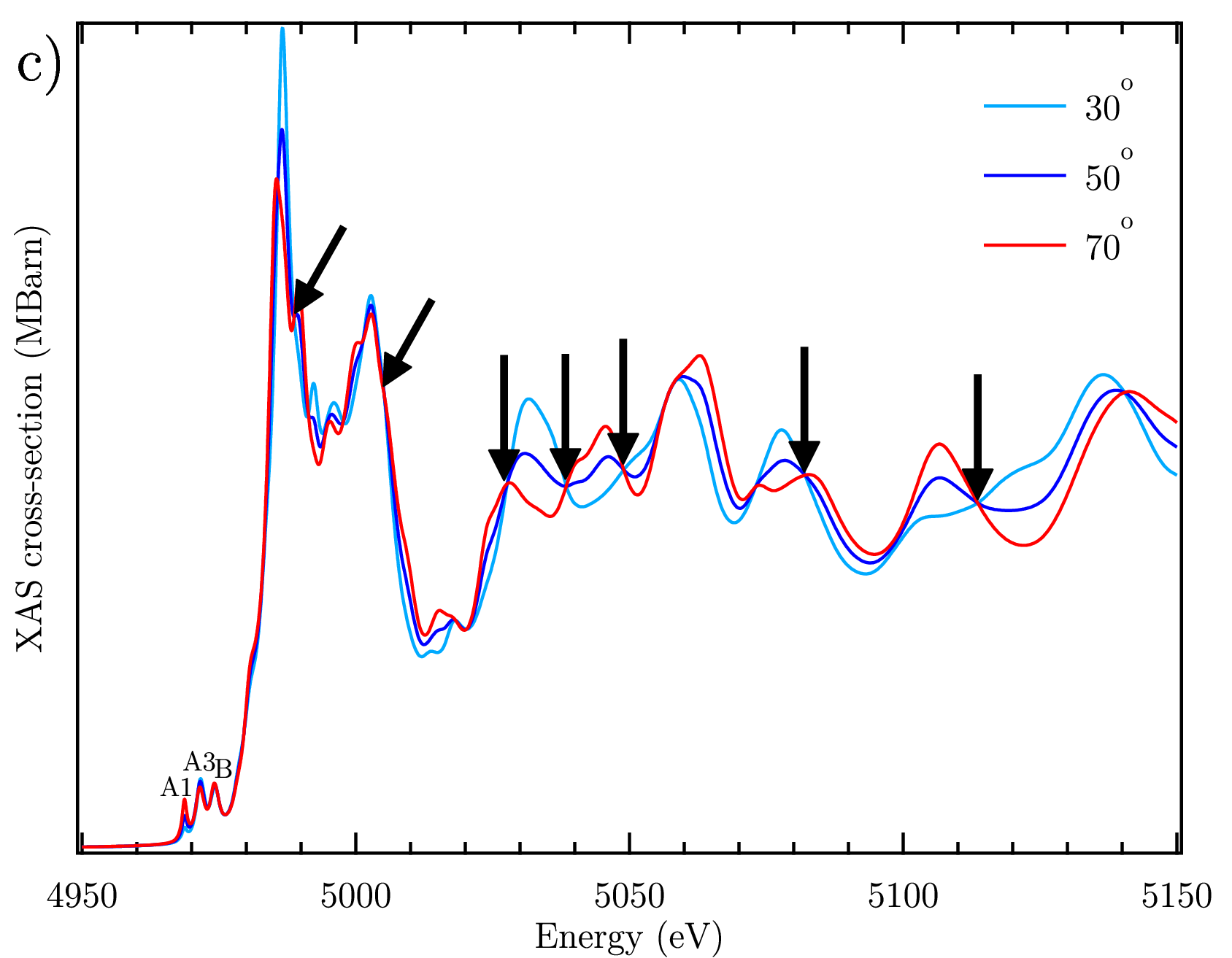}
\includegraphics[scale=0.30,trim={0 0 0 0},clip]{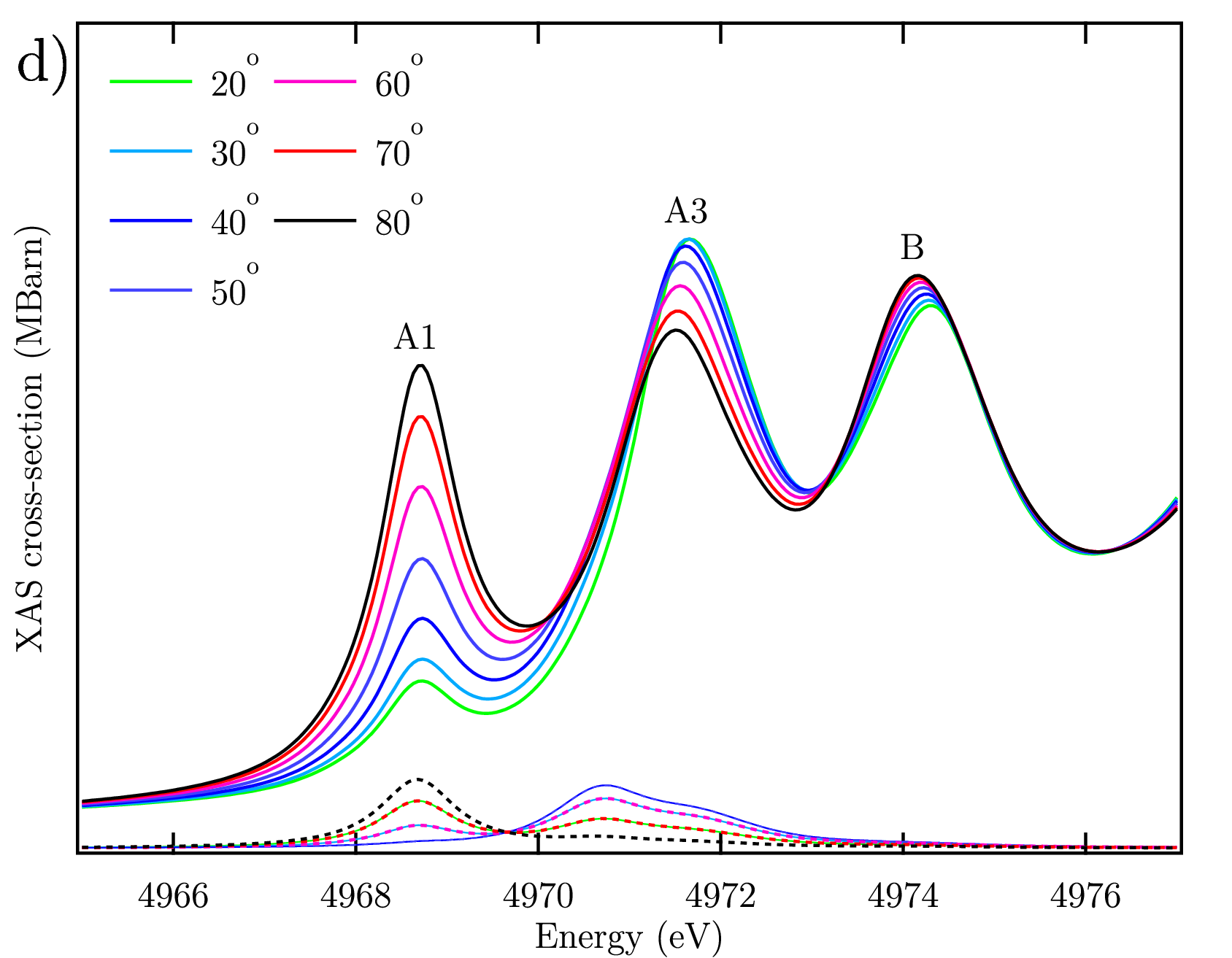}
\end{center}
\caption{a), b) Experimental and c), d) calculated XAS spectra at the \ch{Ti} K-edge of a-\ch{TiO2} for different incidence $\theta$ angles with the sum of dipolar and quadrupolar components (thick lines) and with quadrupolar components only (thin lines and dashed lines in d) for better visibility of overlapping curves). The full XAS is shown in a), c) while the pre-edge is shown in b), d). A few points with $\theta$-independent cross-sections are marked with black arrows.}
\label{experiment_and_theory}
\end{figure*}

The experimental evolution of the \ch{Ti} K-edge spectra with $\theta$ is depicted in Figure \ref{experiment_and_theory}a,b. The spectra are normalized at \SI{4988.5}{\electronvolt} where the cross-section is expected to be $\theta$-independent according to \ac{FDMNES} calculations (shown by the leftmost black arrow in Figure \ref{experiment_and_theory}c). From this normalization procedure, a series of energy points with cross-section independent of the $\theta$ angle appear in the experimental dataset, as predicted by the theory (black arrows in Figure \ref{experiment_and_theory}a and \ref{experiment_and_theory}c) showing the reliability of the normalization procedure. In the pre-edge, the amplitude of peak A1 is dramatically affected by the sample orientation. In the post-edge regions, significant changes are observed as well.

\begin{figure}
\includegraphics[scale=0.50,trim={0 0 0 0},clip]{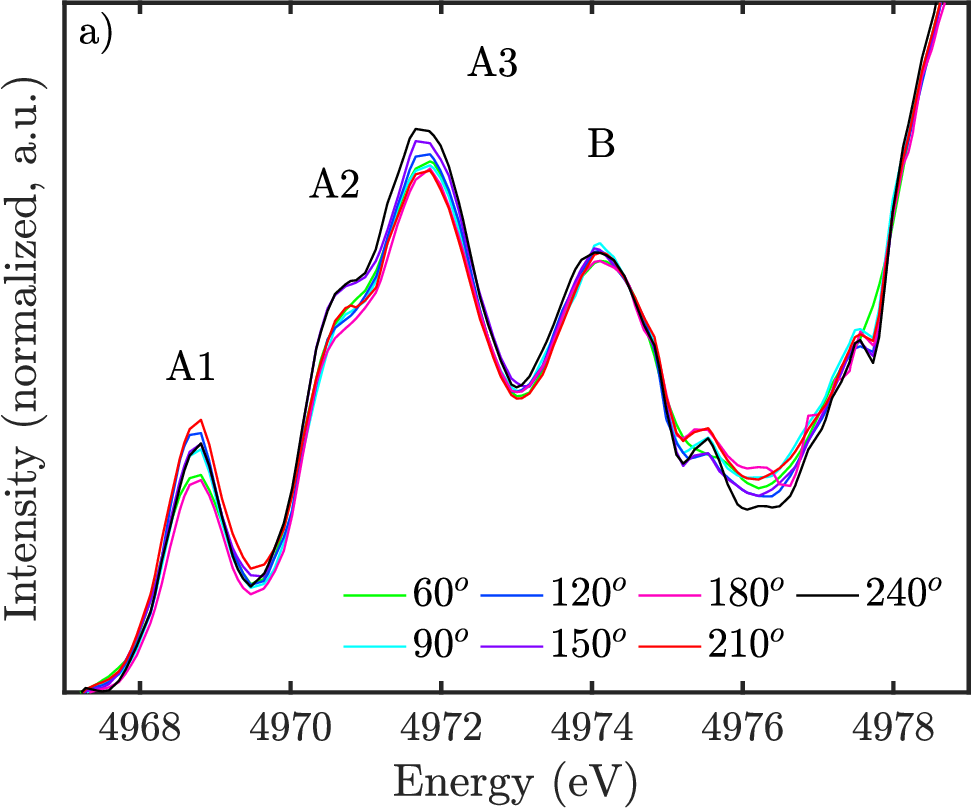}
\includegraphics[scale=0.50,trim={0 0 0 0},clip]{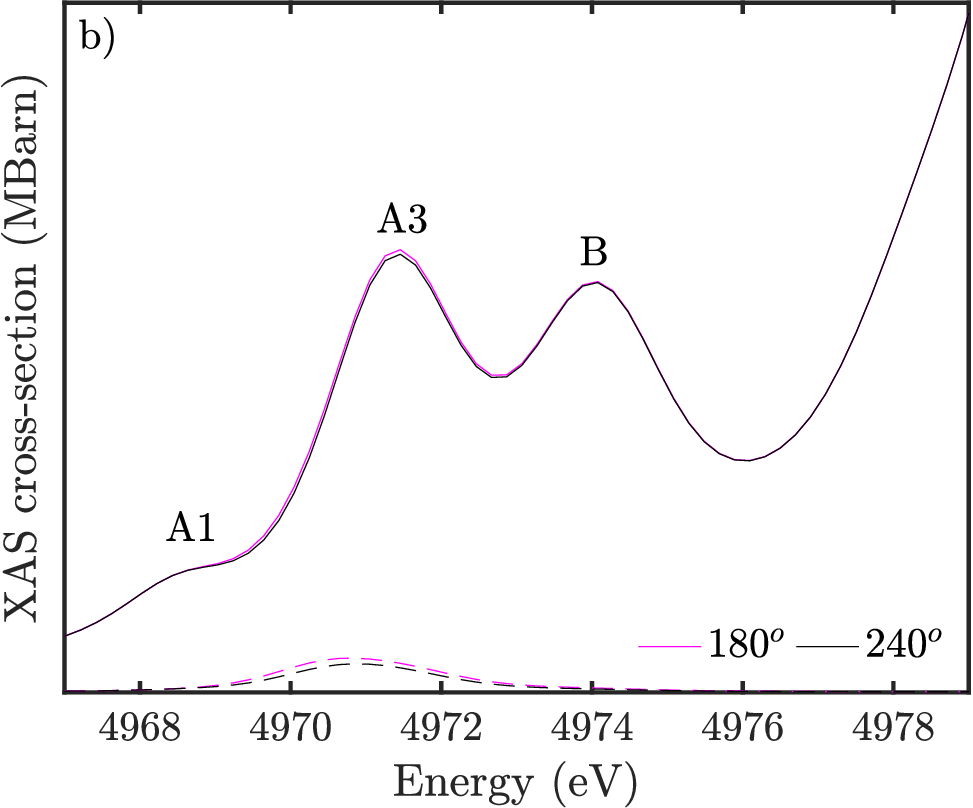}
\caption{a) Experimental and b) calculated XAS spectra at the \ch{Ti} K-edge of a-\ch{TiO2} for different sample orientations along $\phi$ ($\theta=\SI{45}{\degree}$) with the sum of dipolar and quadrupolar components (thin lines) and with quadrupolar components only (thin dashed lines in b)). Only two spectra are shown in b) because only two different spectra can be sampled in \SI{30}{\degree} steps due to the periodicity of \SI{90}{\degree} of quadrupolar transitions.}
\label{experiment_and_theory_mu}
\end{figure}

\emph{Ab-initio} \ac{FDM} calculations of the total \ac{XAS} cross-section (including dipolar and quadrupolar terms) are presented in Figure \ref{experiment_and_theory}c,d for the same angles of incidence $\theta$ as in the experiment. In the pre-edge region, the trends for peak A1 and A3 are nicely reproduced. The absence of peak A2 at first sight, partially originating from defects \cite{Luca:2009bj,Chen:1997by,Luca:1998dn,Hanley:2002go}, is due to our perfect crystal modelling in the \ac{FDM} calculations. In the post-edge region, a good agreement is found, especially for the isosbestic points. This shows that a strong \ac{LD} remains well above the edge in this material. 


The evolution of the spectra is also shown for a fixed incidence angle $\theta=\SI{45}{\degree}$ while the sample is rotated around $\phi$ (Figure \ref{experiment_and_theory_mu}a)\footnote{The normalization energy is at \SI{4988.5}{\electronvolt}}. The changes in amplitude are significantly less than with $\theta$-rotation. We observe a minimal evolution of the amplitudes of peak B and at the rising edge from \SI{4971}{\electronvolt} while a larger effect is distinguished in the spectral region of peaks A1, A2 and A3. \emph{Ab-initio} calculations with the same $\hat \epsilon$ and $\hat k$ orientations as in the experiment are depicted in Figure \ref{experiment_and_theory_mu}b. Only a weak evolution of the amplitude of the pre-edge features is expected and is essentially located in the region of peaks A2 and A3. The amplitude should reach its maximum for $\phi=\SI{180}{\degree}[\SI{90}{\degree}]$ which is inconsistent with the experiment. Instead the fitted evolution of the pre-edge peak amplitudes shows that A2 undergoes a 30\% peak amplitude change whose angular variation is compatible with a quadrupolar transition (\ac{SI} Figure 6a \cite{Note1}) while A1, A3 and B have a maximum amplitude evolution of 10\% (within the fitting confidence interval) with no specific periodicity (\ac{SI} Figure 7 \cite{Note1}). The strong variation in A2 peak amplitude can be observed by the appearance of a pronounced shoulder for $\phi=\SI{150}{\degree}$ which becomes smoother for $\phi=\SI{180}{\degree}$ (Figure \ref{experiment_and_theory_mu}a). Consequently, the main evolution in the pre-edge under $\phi$-rotation is due to peak A2 which explains the disagreement with the perfect crystal \ac{FDM} calculations. It also shows the essentially dipolar content of peaks A1, A3 and B which provide circles in polar plots along $\phi$ (\ac{SI} Figure 7 \cite{Note1}) which is in agreement with the results obtained from $\theta$-scans (\emph{vide infra}). A fit of the A2 peak with a \SI{90}{\degree}-periodic function shows that it may be assigned to the contribution of $d_{x^2-y^2}$ orbitals from the expected angular evolution by spherical harmonic analysis (\ac{SI} Figure 6a,b \cite{Note1}). However, the $d_{x^2-y^2}$ \ac{DOS} in the region of peak A2 is negligible with respect to $d_{xy}$ and $d_{z^2}$ (\emph{vide supra}) hence we rely on the more pronounced angular evolution with $\theta$ in the following to show the involvement of $d_{xy}$ orbitals in the formation of peak A2.

\begin{figure}
\begin{center}
\includegraphics[scale=0.23,trim={0 0 0 0},clip]{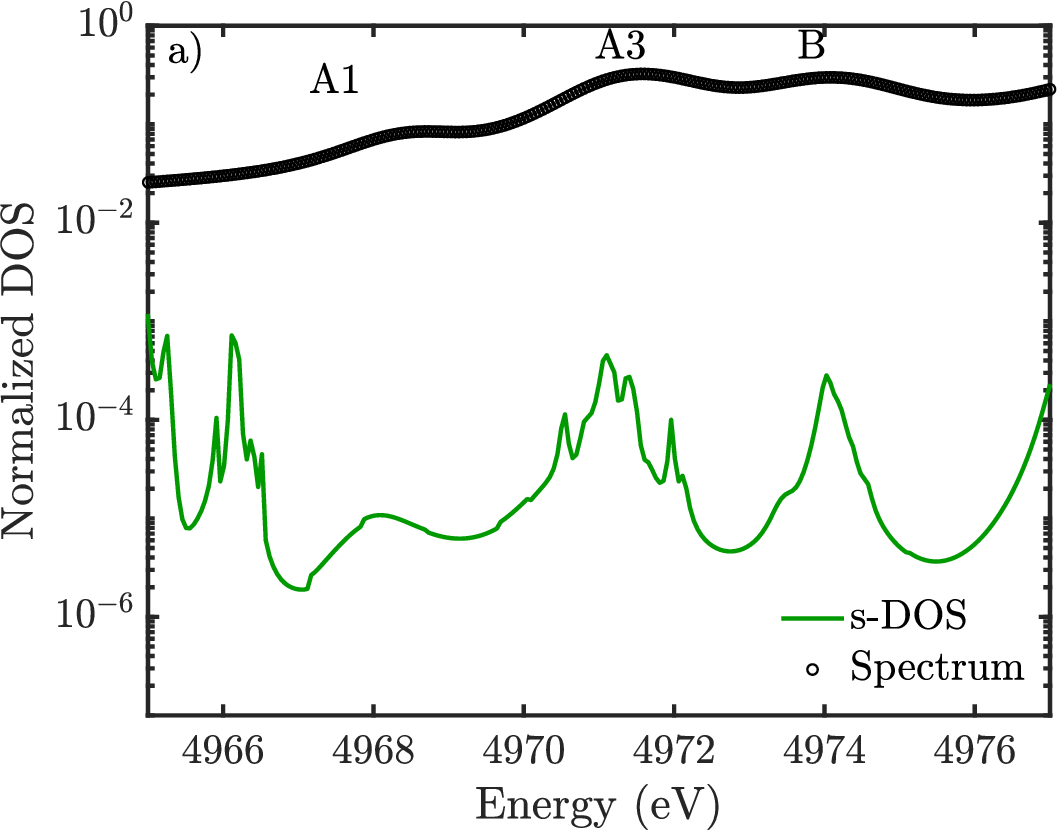}
\includegraphics[scale=0.23,trim={0 0 0 0},clip]{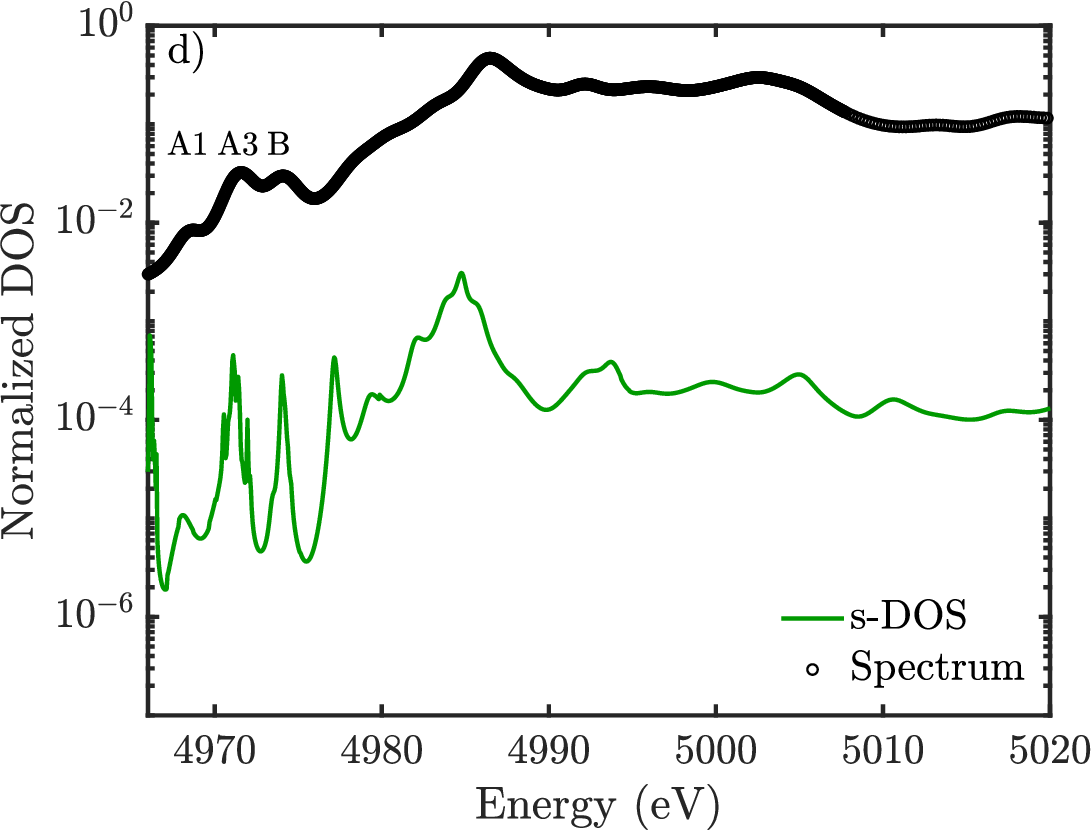} \\
\includegraphics[scale=0.23,trim={0 0 0 0},clip]{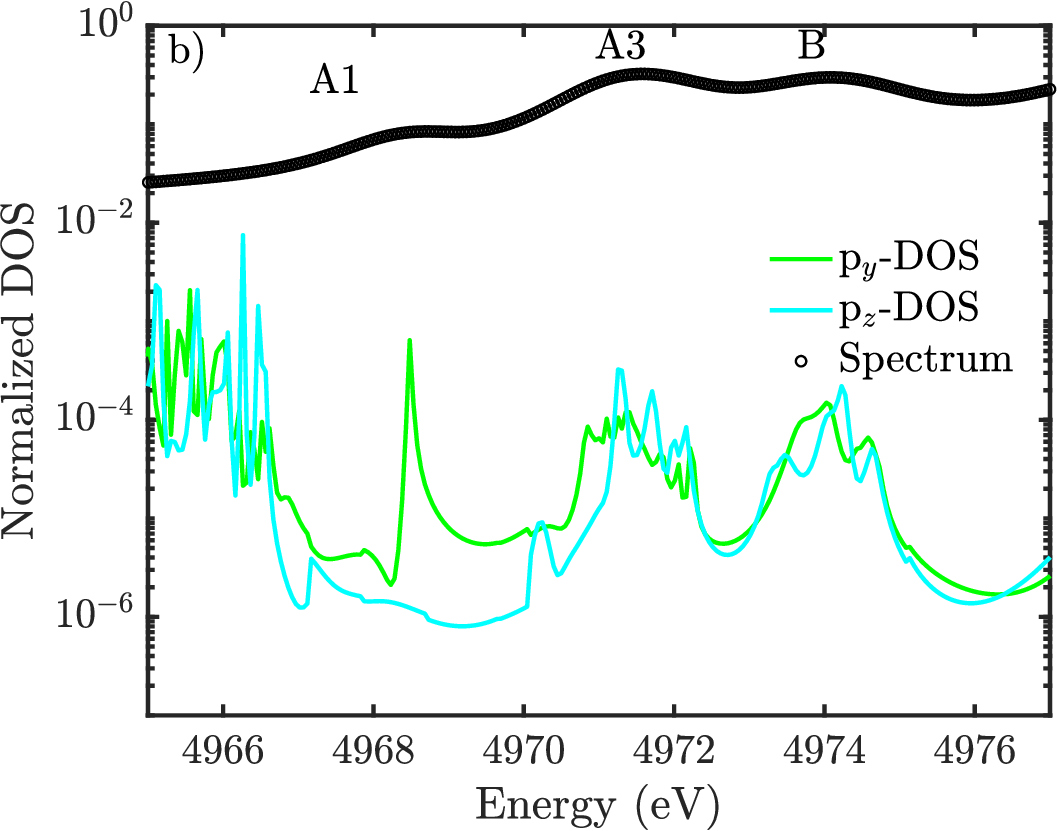}
\includegraphics[scale=0.23,trim={0 0 0 0},clip]{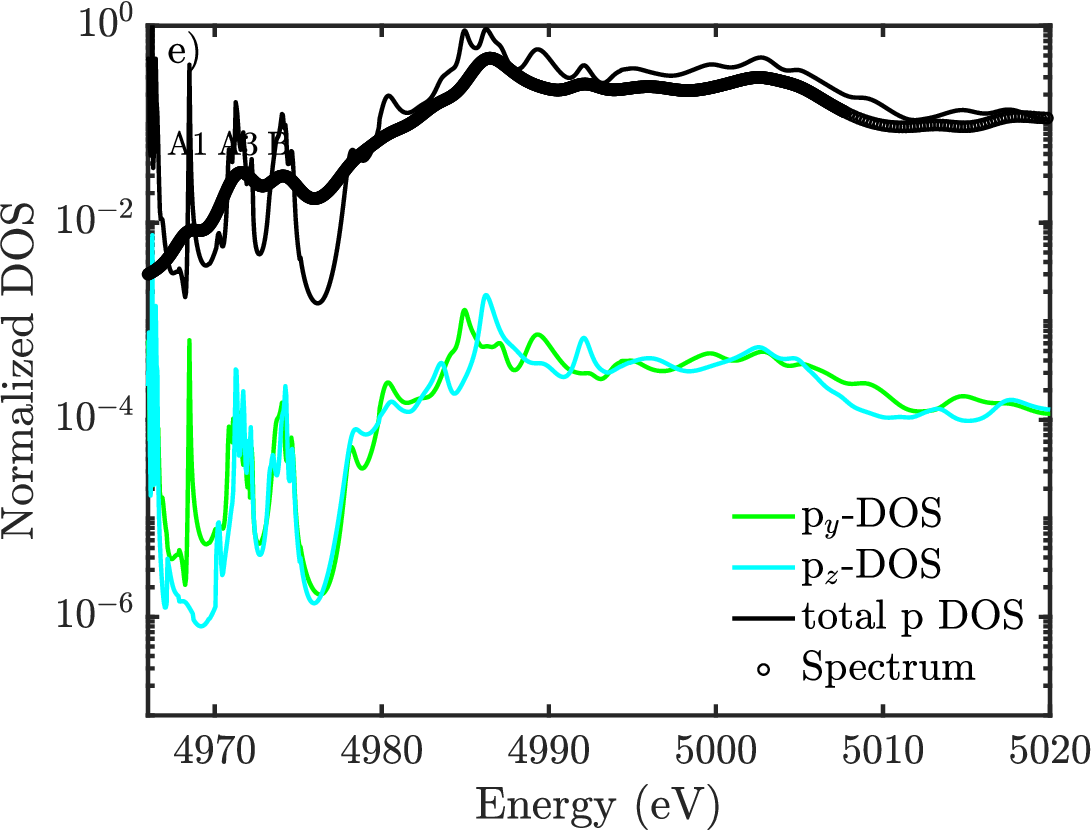} \\
\includegraphics[scale=0.23,trim={0 0 0 0},clip]{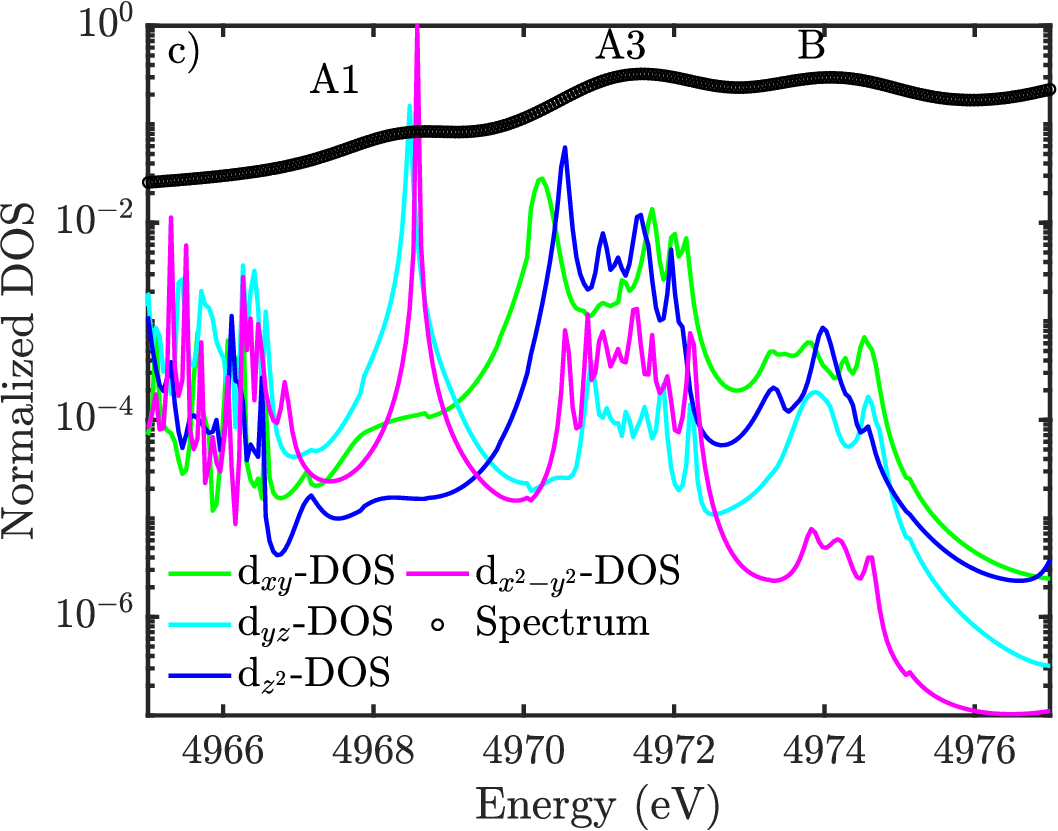}
\includegraphics[scale=0.23,trim={0 0 0 0},clip]{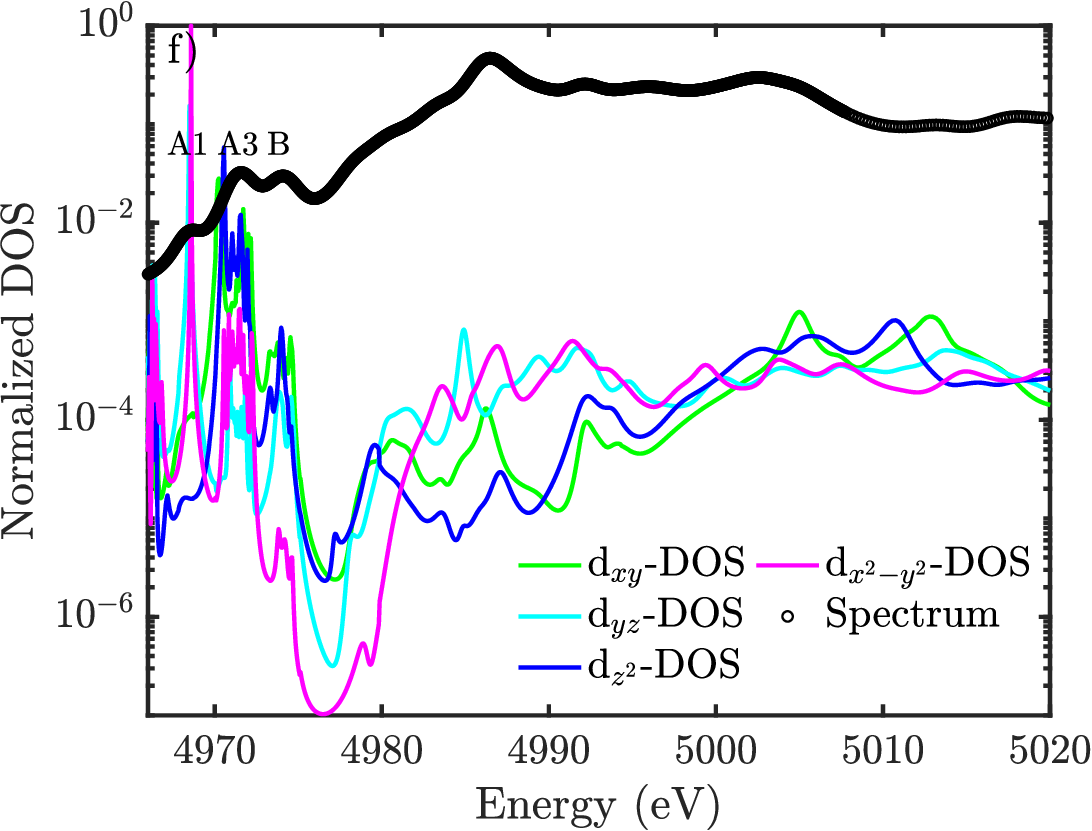} \\
\end{center}

\caption{Calculated projected final state \ac{DOS} for each type of (a,d) $s$-, (b,e) $p$- and (c,f) $d$-final state orbitals in the pre-edge (left) and post-edge regions (right). Reported spectra (black circles) are calculated for $\theta=\SI{90}{\degree}$. The total $p$-\ac{DOS} $(p_x,p_y,p_z)$ is given in e) (black line).}
\label{DOS_anatase}
\end{figure}

In order to describe the origin of the \ac{LD} with $\theta$ and assign the pre-edge resonances, the projected \ac{DOS} of the final states for the pre-edge and post-edge region is depicted in Figure \ref{DOS_anatase} (we drop the term "projected" in the following for simplicity). Due to the large differences between the \ac{DOS} of $s$, $p$ and $d$ states, a logarithmic scale is used vertically and normalized to the orbital having the largest \ac{DOS} contributing to the final state among $s$, $p$ and $d$ orbitals. For peaks A1, A3 and B, most of the \ac{DOS} comes from $d$-orbitals while $s$- and $p$-\ac{DOS} are comparable. However, due to the angular momentum selection rule, the spectrum resembles the $p$-\ac{DOS} both in the pre-edge and the post-edge regions as witnessed by the similarity between the total $p$-\ac{DOS} and the calculated spectrum (black line in Figure \ref{DOS_anatase}e). Importantly, peak A1 has only ($p_x,p_y$) contributions (Figure \ref{DOS_anatase}b), meaning that this transition is expected to have a much weaker intensity when the electric field gets parallel to the $\hat z$ axis, in agreement with the $\theta$-dependence of its amplitude (Figure \ref{experiment_and_theory}). The $d$-\ac{DOS} at peak A1 involves d$_{xz}$, d$_{yz}$ and d$_{x^2-y^2}$ orbitals (Figure \ref{DOS_anatase}c), among which the first two can hybridize with the ($p_x$,$p_y$) orbitals and relax the dipole selection rules. The dipolar nature of A1 is also seen from the monotonic increase of its amplitude from $\theta=\SI{0}{\degree}$ to $\theta=\SI{90}{\degree}$, inconsistent with a quadrupolar allowed transition with \SI{90}{\degree} periodicity. Following the same analysis, peaks A3 and B do not undergo a strong change in amplitude under $\theta$-rotation because $(p_x,p_y)$ and $p_z$ contribute similarly to the \ac{DOS} for these transitions (Figure \ref{DOS_anatase}b) although \ac{FDM} calculations show that A3 should evolve in intensity with $\theta$ due to a $\sim$20\% larger \ac{DOS} for $p_z$ than for $p_x,p_y$ as experimentally observed. From the integrated $d$-\ac{DOS} along $(x,y)$ and $z$ (\ac{SI} Figure 8a \cite{Note1}), we notice the inconsistency between the peak amplitudes in the theory and the experiment, which shows that they are essentially determined by the $p$-\ac{DOS} (\ac{SI} Figure 8b \cite{Note1}).

\begin{figure}
\begin{center}
\includegraphics[scale=0.24]{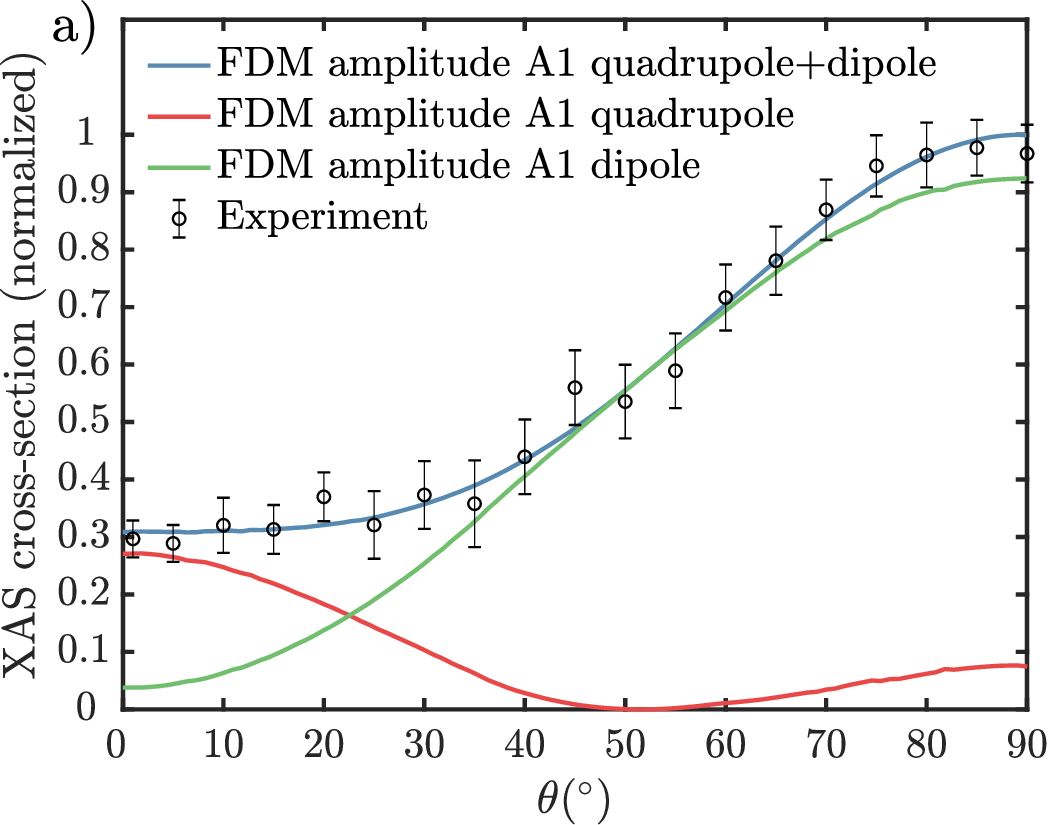}
\includegraphics[scale=0.24]{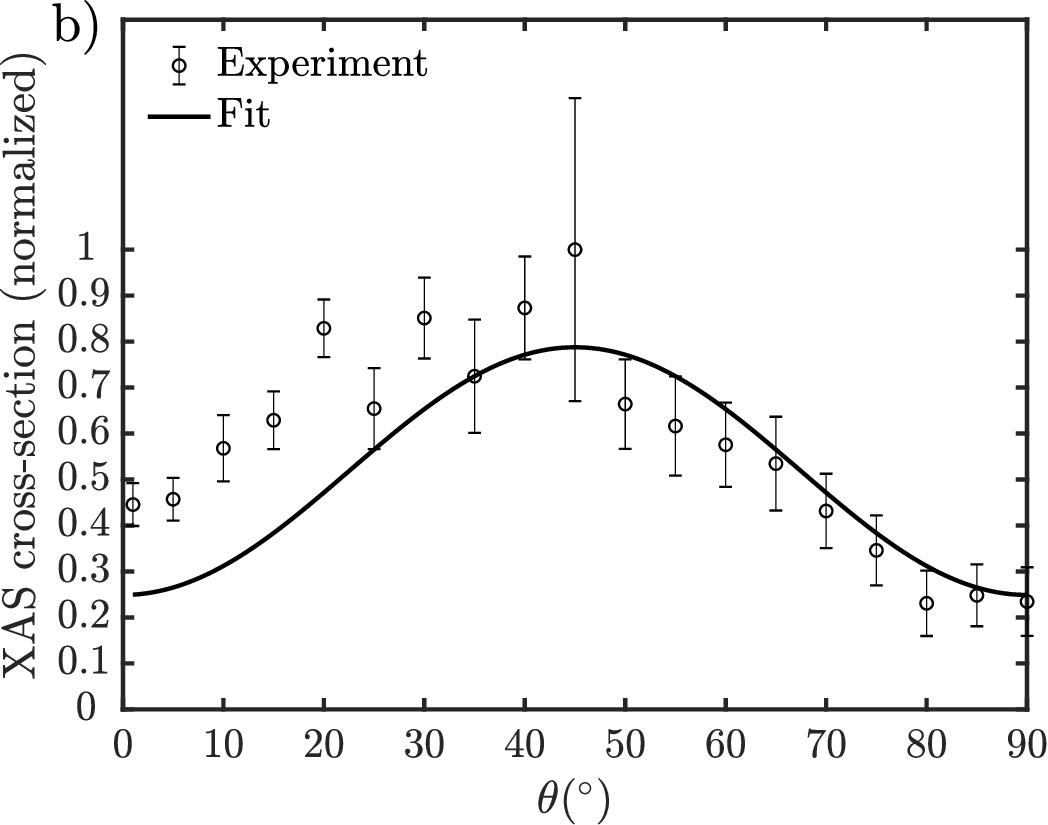}
\caption{Overlap between the angular evolution of the amplitudes for peaks a) A1, b) A2 in the theory (lines) and the experiment (circles with error bars). The error bars represent 95\% of confidence interval for the fitting of the amplitude.}
\label{fitting_pre_edge}
\end{center}
\end{figure}

For a more quantitative description of the dipolar and quadrupolar components in the pre-edge, we extracted the quadrupolar cross-section from \ac{FDM} calculations. It is depicted as thin lines in Figure \ref{experiment_and_theory}d where continuous and dashed lines are used for better visibility of overlapping curves. The quadrupolar contributions are limited to peaks A1 and A3 with a contribution in the spectral region of peak A2. At peak A1, the quadrupolar amplitude is maximum for $\theta=\SI{0}{\degree}$ and $\theta=\SI{90}{\degree}$ and the total cross-section becomes mainly quadrupolar for $\theta=\SI{0}{\degree}$ while the quadrupolar component contributes $\sim\SI{15}{\percent}$ of the peak amplitude for $\theta=\SI{90}{\degree}$. From the development of the cross-section into spherical harmonics (Table \ref{dipole_table}), the dipolar transitions to $p_{x,y}$ final states are expected to vary as $-\cos^2\theta$ while transitions to $p_z$ vary as $\cos^2\theta$ plus a constant (see \ac{SI} Figure 12a \cite{Note1}). The fitted evolution of the dipolar cross-section of peak A1 in the experiment and in the \ac{FDM} calculations is compatible with a transition to $p_{x,y}$ (green line in Figure \ref{fitting_pre_edge}a, fitting details in \ac{SI} \S2 \cite{Note1}). The quadrupolar component (red line in Figure \ref{fitting_pre_edge}a) is compatible with a transition to $d_{xz},d_{yz}$ due to its $-\sin^2\theta\cos^2\theta$ predicted evolution (Table \ref{dipole_table}) in agreement with the $d$-\ac{DOS} at peak A1 (Figure \ref{DOS_anatase}c and Figure 12b in \ac{SI}) \cite{Note1}. The comparison between the experimental and theoretical amplitudes of peak A1 (Figure \ref{fitting_pre_edge}a) gives an excellent agreement further confirming that the A1 transition is mostly dipolar to $p_{x,y}$ final states. Following the same analysis, it is more difficult to determine the dominant $p$-\ac{DOS} contributing to the transitions at peak A3 and B due to the weak evolution of their amplitude with $\theta$. 

As pointed out earlier, the quadrupolar cross-section has a doublet structure in the region of peaks A2 and A3 (Figure \ref{experiment_and_theory}d). The most intense of the two peaks at $\theta=\SI{45}{\degree}$ is in the spectral region of peak A2 where the transition involving defects is expected in a-\ch{TiO2} nanoparticles for instance. A closer look at the fitted evolution of the A2 amplitude with $\theta$ shows a quadrupolar evolution with maximum value at $\theta=\SI{45}{\degree}$ (Figure \ref{fitting_pre_edge}b). This is in agreement with the expected angular evolution of $d_{z^2}$ and $d_{xy}$ final states from spherical tensor analysis (\ac{SI} Figure 12b \cite{Note1}) which contribute to the \ac{DOS} in the spectral region of peak A2 (Figure \ref{DOS_anatase}c). It indicates that although the amplitude of A2 is underestimated in the \ac{FDM} calculation, the consensus that A2 originates from undercoordinated and disordered samples may be more subtle because of the involvement of a transition in the perfect crystal and is discussed in the next section.

From this combined experimental and theoretical analysis, we emphasize that consecutive peaks in the pre-edge of a-\ch{TiO2} are not simply due to the energy splitting between t$_{2g}$ and e$_g$ as previously invoked \cite{Cabaret:2010fp}. This splitting is more complicated than the usual octahedral crystal field splitting because of the strong hybridization between $p$ and $d$ orbitals in a lowered symmetry environment which affects the relative ordering of the transitions. The consistent results between experiment, \ac{FDM} calculations and spherical tensor analysis show the reliability of the assignment provided in this work. Table \ref{literature_assignment} compares our results with previous assignments of peaks A1 to B.

\section{Discussion}


\subsection{Local versus non-local character of the pre-edge transitions}

\begin{figure}
\includegraphics[scale=0.5]{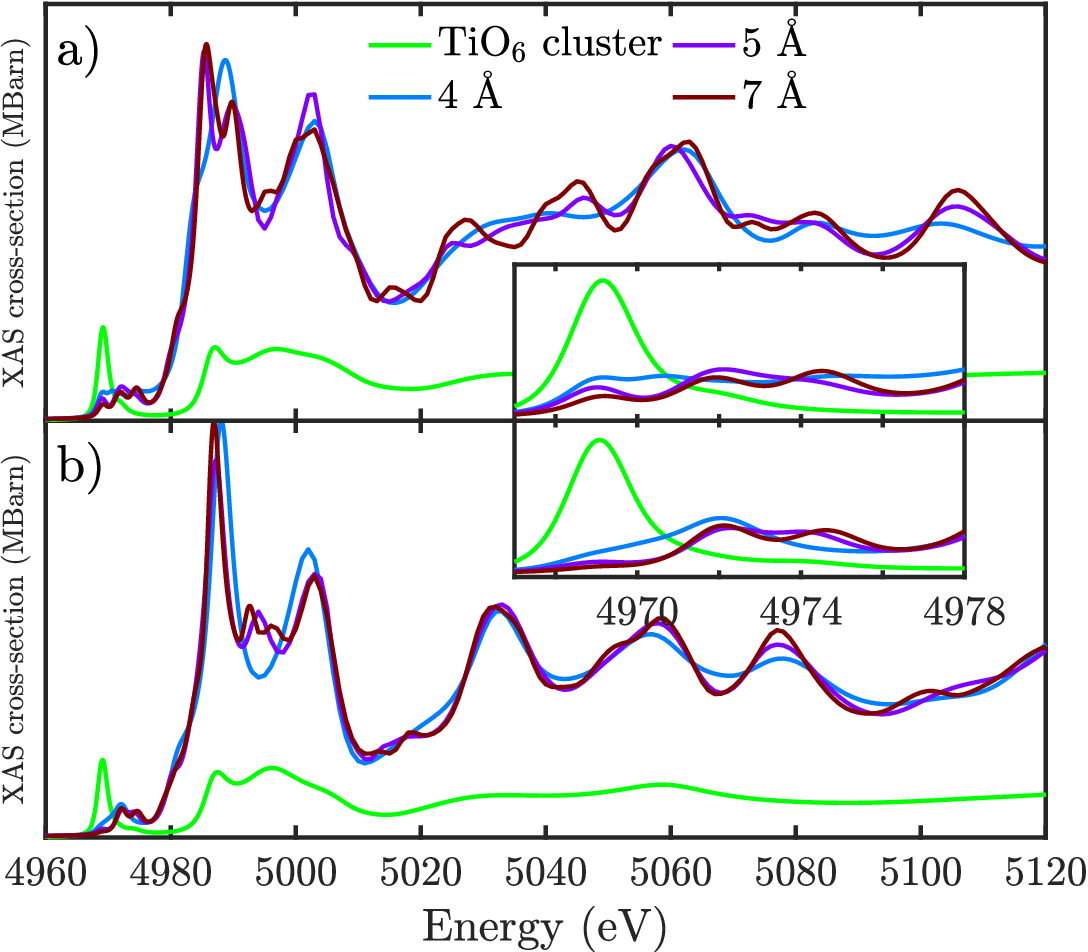}
\caption{Evolution of the calculated a-\ch{TiO2} XAS spectrum at the \ch{Ti} K-edge with cluster size in the \ac{FDM} calculation for a) $\theta=\SI{90}{\degree}$ ($\hat\epsilon\parallel[010]$,$\hat k\parallel[00-1]$), b) $\theta=\SI{0}{\degree}$ ($\hat\epsilon\parallel[001]$,$\hat k\parallel[100]$).}
\label{anatase_size_convergence}
\end{figure}

Pre-edge transitions can originate either from on-site (localized) or off-site transitions involving neighbour \ch{Ti} atoms of the absorber. Off-site transitions are dipole allowed due to the strong $p-d$ orbital hybridization \cite{Yamamoto:2008bi}. This effect has been shown on \ch{NiO}, an \ac{AF} charge-transfer insulator, for which the \ch{Ni} K-edge transition to $3d$ orbitals of the majority spin of the absorber is only possible between \ch{Ni} sites due to the \ac{AF} ordering \cite{Gougoussis:2009kr}. Hence, to disentangle between the local or non-local character of the pre-edge transitions in a-\ch{TiO2}, we performed \ac{FDM} calculations on clusters with increasing number of neighbour shells starting from an octahedral \ch{TiO_6} cluster with the same geometry and bond distances as in the bulk. The results are shown in Figure \ref{anatase_size_convergence} with two orthogonal electric field orientations along $[001]$ ($\theta=\SI{0}{\degree}$) and $[010]$ ($\theta=\SI{90}{\degree}$).

The calculation for \ch{TiO6} (green curve) shows only A1 meaning that it is mostly an on-site transition. The absence of peaks A3 and B suggests that they are mostly non-local transitions in agreement with Ref.\ \cite{Cabaret:2010fp}. Increasing the cluster size to \SI{4}{\angstrom} includes the second shell of \ch{Ti} ions, which generates most of the A3 amplitude. This shows that, similarly to \ch{NiO}, an energy gap opens between the on-site and off-site transitions to $3d$ orbitals of \ch{Ti} and that A3 is mostly dipolar and strongly influenced by the intersite $3d$--$4p$ hybridization. Peak B is missing for this cluster size which shows that it is due to a longer range interaction and can be reconstructed with a \SI{5}{\angstrom} cluster including the next shell of neighbor \ch{Ti} atoms. A similar trend in the local or non-local character of the pre-edge transitions is observed at the metal K-edge of $3d$ transition metal oxides which has to do with the degree of $p$--$d$ orbital mixing \cite{Wu:2004bs}.


\subsection{Origin of peak A2 in bulk a-\ch{TiO2}}

\label{A2_origin}

The experimental $\phi$ and $\theta$ angular evolution of the A2 amplitude (Figure \ref{fitting_pre_edge}b and 6a in the \ac{SI} \cite{Note1}) matches a quadrupolar transition, qualitatively consistent with the dominant quadrupolar cross-section obtained from \ac{FDMNES} calculations in the region of peak A2 (thin lines in Figure \ref{experiment_and_theory}d and Figure \ref{experiment_and_theory_mu}b). Recent calculations accounting for the electron-hole interaction in the Bethe-Salpeter equation have reproduced peak A2, although with an underestimated amplitude as in our \ac{FDM} calculations \cite{Vorwerk:2017gs}. Peaks A1 and A2 are found to exhibit their maximum amplitude when the electric field is parallel to the $(\textbf{a},\textbf{b})$ and $\textbf{c}$ axes, respectively. This is in agreement with our measurement for peak A1 (Figure \ref{fitting_pre_edge}a) as a result of the coupling between the $3d$ states of \ch{Ti} with the $p_{x,y}$ \ac{DOS}. For peak A2 (Figure \ref{fitting_pre_edge}b), we observe a dominant quadrupolar evolution with a deviation from the ideal behavior showing the presence of $p_z$ states which increase their contribution to the transition when $\theta\rightarrow\SI{0}{\degree}$ (see blue line in Figure 12a \ac{SI} \cite{Note1}). Although both peaks A1 and A2 show \emph{p-d} orbital mixing, this mixing is clearly stronger for the A1 peak where the dipole contribution becomes dominant over the quadrupolar in contrast to the A2 peak. It shows that the amount of \emph{p-d} orbital mixing differs for these transitions which can be explained by the $\sim$100 times lower $p_z$ \ac{DOS} in the region of peak A2 than $p_{x,y}$ \ac{DOS} in the region of peak A1 (Figure \ref{DOS_anatase}b). The underestimated amplitude of peak A2 in our calculation is likely due to the lack of explicit treatment of the electron-hole interaction which would improve the agreement of energy and amplitudes for peaks A1 and A2 without resorting to changes in screening constants of the $3d$ electrons as in our study. Recent calculations using the Bethe-Salpeter equation estimate that the average amplitude of peak A2 is $\sim\SI{15}{\percent}$ of the average amplitude of peak A1 which comforts this hypothesis \cite{Vorwerk:2017gs}. A parallel can be made between the energy splitting of peaks A1 and A2 containing quadrupolar localized components and the splitting of the bound excitons of a-\ch{TiO2} observed in the optical range where the $(\textbf{a},\textbf{b})$ plane exciton has a larger binding energy than the $\textbf{c}$ exciton \cite{Chiodo:2010eh,Kang:2010bj,Baldini:2017kv}.

\subsection{Origin of peak A2 in defect rich a-\ch{TiO2}}

\begin{figure}
\includegraphics[scale=0.25]{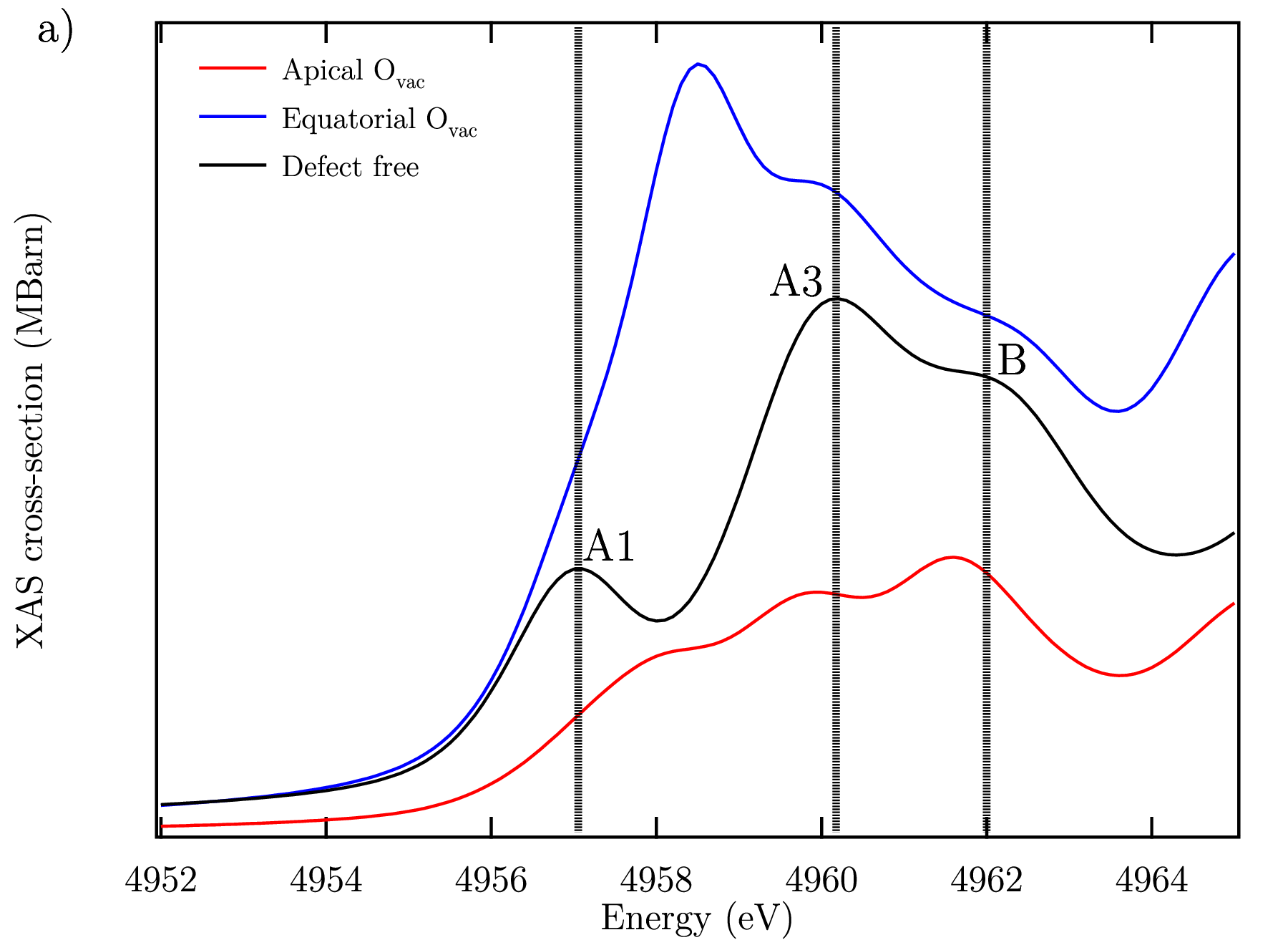}
\includegraphics[scale=0.25]{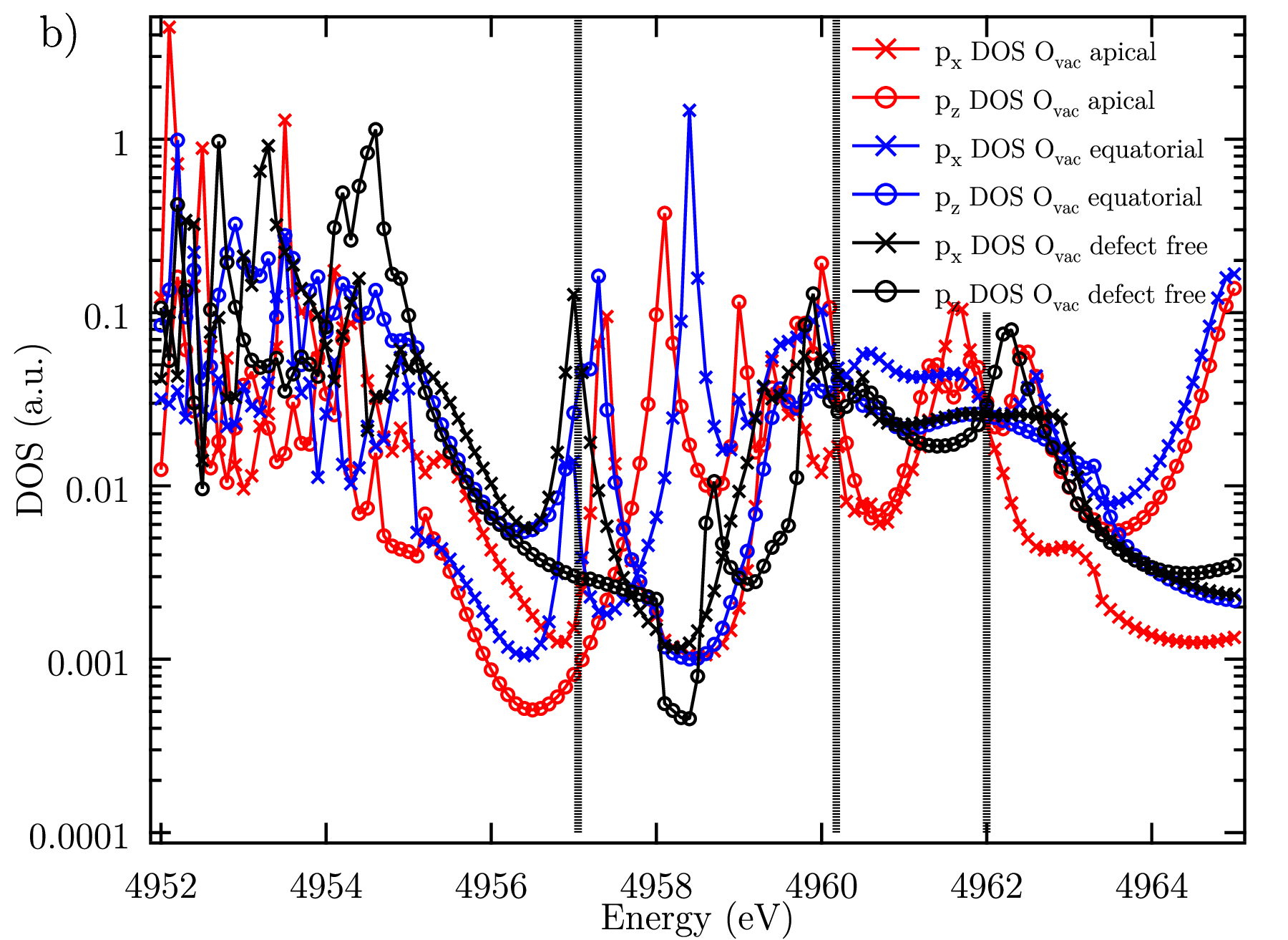}
\caption{a) Effect of an oxygen vacancy introduced at the equatorial (blue) or apical position (red) of a \ch{TiO_6} octahedron on the \ac{XAS} spectrum of a $4\times4\times4$ a-\ch{TiO2} supercell including lattice relaxation. The spectrum at the \ch{Ti} K-edge for the perfect supercell is shown in black. The calculation is angle averaged (no specific orientation taken for the crystal and the incident X-ray beam). b) Computed $p$-\ac{DOS} for defect free a-\ch{TiO2} (black) and with an oxygen vacancy in the equatorial (blue) and apical position (red). The $p_{x,y}$ \ac{DOS} is shown with crosses and the $p_z$ \ac{DOS} with circles. Vertical dotted lines are guides to the eye at the energy position of the peaks maxima in the pre-edge of defect free a-\ch{TiO2}.}
\label{defect_simulation}
\end{figure}

While we show that the presence of the A2 peak can be explained by the electronic structure of a-\ch{TiO2}, a number of previous studies have concluded that A2 is related to lattice defects \cite{Luca:2009bj,Chen:1997by,Luca:1998dn,Hanley:2002go,RittmannFrank:2014fu,Santomauro:2016bg,Budarz:2017iu}. The question arises as to the connection between the A2 peak and the lattice defects, if any. Oxygen vacancies are native defects in a-\ch{TiO2} \cite{Morgan:2009cd}. The occurrence of an oxygen vacancy (\ch{O_{vac}}) in the vicinity of a \ch{Ti} atom will further lower the D$_{2d}$ symmetry and introduce $p-d$ orbital mixing in the pentacoordinated \ch{Ti} atom increasing the transition amplitude while broadening the transition due to the inhomogeneous contribution of the vacancy distribution \cite{Zhang:2008gt,Triana:2016fi}. In order to check the effect of an \ch{O_{vac}} on the \ac{XAS} spectrum of a \ch{Ti} atom in the vicinity, \emph{ab initio} \ac{FDM} calculations are performed at the \ch{Ti} K-edge of \ch{Ti} atoms with a doubly ionized \ch{O_{vac}} (\ch{V_O^{2+}}) at the apical or equatorial position in a supercell of 768 atoms. The calculations are performed with a bulk a-\ch{TiO2} $4\times4\times4$ superlattice structure from which one oxygen atom is removed in the center and neighbor titanium atoms are moved along the broken \ch{Ti}--\ch{O} bond to simulate lattice relaxation. We have taken the local structural relaxation reported in another work with hybrid functional calculations where the titanium atoms move away from \ch{V_O^{2+}} by \SI{0.509}{\angstrom} in the equatorial plane and \SI{0.109}{\angstrom} in the apical position \cite{Finazzi:2008ff}. The results are depicted in Figure \ref{defect_simulation}a for the absorption cross-section and \ref{defect_simulation}b for the \ac{DOS} \footnote{The energy axis of Figure \ref{defect_simulation} is shifted with respect to the other energy scales reported in the manuscript by $\sim\SI{12}{\electronvolt}$. This is because the FDMNES calculations provide a calculated spectrum which is energy shifted with respect to the experiment. A shift is applied for a straightforward comparison between calculations and experiment. This energy shift is not applied in Figure \ref{defect_simulation}. Calculations performed with the space group or with the supercell for a random polarization are shown in the Supplementary Material Figure 11 \cite{Note1}.}. The calculations with an oxygen vacancy show a peak in the region between peak A1 and peak A3 where peak A2 is expected while peak A3 and B remain essentially unaffected. The calculated \ac{DOS} (Figure \ref{defect_simulation}b) shows that the $p_{x,y}$ \ac{DOS} at peak A1 is slightly blue shifted when the vacancy is introduced in the apical position while for the vacancy in the equatorial position, the $p_z$ \ac{DOS} dominates in the region of peak A1 with a slight blue shift. The calculated \ac{DOS} in the regions of peaks A3 and B shows essentially changes in the amplitudes with almost no chemical shift except for a substantial red shift of the $p_{x,y}$ \ac{DOS} of peak B when the vacancy is introduced at the apical position. In contrast to these minor changes along the energy axis, the largest difference appears in the energy region between peaks A1 and A3 where a $p_{x,y}$ and $p_{z}$ \ac{DOS} appears upon formation of a vacancy in the equatorial and apical position, respectively. Hence, \ac{DOS} is formed in the region of peak A2 which has the polarization corresponding to the orbitals pointing towards the \ch{O_{vac}}. It is clear from this analysis that this \ac{DOS} is representative of the defect states introduced by the \ch{O_{vac}}. We have compared our calculated \ac{DOS} with other authors. Janotti et al.\ report hybrid functionals calculations (HSE) which predict that \ch{V_O^{2+}} forms empty \ac{DOS} \SI{0.65}{\electronvolt} above the \ac{CBM} \cite{Janotti:2010cx} which is comparable to our prediction of \SI{0.71}{\electronvolt} and \SI{1.11}{\electronvolt} for the energy of the calculated \ac{DOS} representative of \ch{V_O^{2+}} with respect to the \ac{CBM}. Na-Phattalung et al.\ find that the unoccupied \ac{DOS} of the nearest titanium atoms to the vacancy has a maximum on the low energy side of peak A3, in agreement with our results for the vacancy in equatorial position \cite{NaPhattalung:2006jh}. Similar results have been obtained in rutile \ch{TiO2} for which doubly ionized oxygen vacancies introduce a blue shift of the \ch{Ti} $3d$-\ac{DOS} by \SI{1}{\electronvolt} \cite{Vasquez:2016jz}. A similar effect is present in a-\ch{TiO2} at the O K-edge where the asymmetry of the so-called $t_{2g}$ and $e_g$ peaks with a tail on the high-energy side cannot be reproduced in the calculations with a bulk structure which is assigned to the presence of blue shifted defect states from the bulk \cite{Kruger:2017bl}. The amplitude of these peaks increases upon heavy ion irradiation compatible with the formation of more oxygen vacancies \cite{Thakur:2011fq}. Hence, we find that the occurrence of a transition corresponding to undercoordinated \ch{Ti} atoms in the region of peak A2 overlapping with the intrinsic quadrupolar transition detailed in \S\ref{A2_origin} is a coincidence. The experimental spectrum of a-\ch{TiO2} nanoparticles with defects would be a linear combination of the \ch{O_{vac}} spectra (red and blue curves in Figure \ref{defect_simulation}) and the spectrum of hexacoordinated \ch{Ti} atoms in the bulk (black curve in Figure \ref{defect_simulation}) which depends on the amount of vacancy in the system. However, this study shows that peak A2 is expected to be present even in crystalline a-\ch{TiO2} nanoparticles because the intrinsic quadrupolar transition is likely dominant over the defect contribution. The large increase in cross-section in the region of peak A2 for pentacoordinated \ch{Ti} atoms is fully compatible with our recent studies on photoexcited a-\ch{TiO2} nanoparticles \cite{RittmannFrank:2014fu,Santomauro:2016bg,Budarz:2017iu}. We therefore conclude that peak A2 originates from an essentially quadrupolar transition in the regular a-\ch{TiO2} lattice and from the $p$-\ac{DOS} of pentacoordinated \ch{Ti} atoms with an \ch{O_{vac}} in the vicinity.

\section{Conclusion}

In summary, a complementary approach using experimental \ac{LD} measurements at the \ch{Ti} K-edge of a-\ch{TiO2}, \emph{ab-initio} \ac{FDM} calculations and spherical tensor analysis provides an unambiguous assignment of the pre-edge features. We show that A1 is mainly due to a dipolar transition to on-site hybridized $4p_{x,y}-3d_{xz},3d_{yz}$ final states which give a strong dipolar \ac{LD} to the transition with a weak quadrupolar component from $(3d_{xz},3d_{yz},3d_{x^2-y^2}$) states. The A3 peak is due to a mixture of dipolar transitions to hybridized $4p_{x,y,z}-(3d_{xy},3d_{z^2})$ final states as a result of strong hybridization with the $3d$ orbitals of the nearest \ch{Ti} neighbour with a small quadrupolar component. The B peak is purely dipolar ($4p$ orbitals in the final state) and is an off-site transition (the electron final state is delocalized around the absorbing atom). The distinction between on-site and off-site transitions is possible using different cluster sizes in the \ac{FDM} calculations. The \ac{LD} is visible well above the absorption edge due to the strong $p$-orbital polarization in a-\ch{TiO2} which affects the amplitude of the \ac{EXAFS}. Surprisingly, a quadrupolar angular evolution of peak A2 is observed for the first time with a narrow bandwidth showing that it is an intrinsic transition of the single crystal. A connection between the unexpectedly large experimental amplitude of this peak in nanoparticles is made with oxygen vacancies forming pentacoordinated \ch{Ti} atoms. Crude FDMNES calculations show that empty \ac{DOS} appears in the region of peak A2 upon formation of oxygen vacancies (\ch{V_O^{2+}}) and that it overlaps with the quadrupolar transition observed in this work for defect free materials. This explains the relatively intense A2 peak in amorphous \ch{TiO2} \cite{Zhang:2008gt} or upon electron trapping at defects after photoexcitation of anatase or rutile \ch{TiO2} \cite{Budarz:2017iu,RittmannFrank:2014fu,Santomauro:2016bg}. The unprecedented quantitative agreement provided in this work is made possible by the continued improvement of computational codes including full potentials \cite{Joly:1999iq,Joly:2001fu,Joly:2009ha} and the more accurate description of the core-hole interaction in Bethe-Salpeter calculations \cite{Shirley:2004id,Vorwerk:2017gs}. Experiments are on-going to extend this work to rutile \ch{TiO2}.

The present results and analysis should be cast in the context of ongoing ultrafast X-ray spectroscopy studies at Free Electron Lasers \cite{Abela:2017jr,Chergui:2016hb}. For materials such as a-\ch{TiO2}, the increased degree of detail that can be gathered from such sources was nicely illustrated in a recent paper by Obara et al.\ \cite{Obara:2017bq} on a-\ch{TiO2}, showing that the temporal response of the pure electronic feature (at the \ch{Ti} K-edge) was much faster ($\sim\SI{100}{\femto\second}$) than the response of structural features ($\sim\SI{330}{\femto\second}$) such as the pre-edge and the above-edge \ac{XANES}. The present work shows that by exploiting the angular dependence of some of the features, even up to the \ac{EXAFS} region, one could get finer details about the structural dynamics, in particular, of non equivalent displacements of nearest neighbours.


\begin{acknowledgments}
We thank Yves Joly and Christian Brouder for fruitful discussions and Hengzhong Zhang for providing the \ac{FDMNES} input files. We also thank Beat Meyer and Mario Birri of the microXAS beamline for their technical support as well as the Bernina station staff of the SwissFEL for lending us the goniometer stage. This work was supported by the SNSF via the NCCR:MUST and grants 200020\_169914 and 200021\_175649 and the European Research Council (ERC) Advanced Grants H2020 ERCEA 695197 DYNAMOX. G.\ F.\ M.\ and C.\ B.\ were supported via the InterMUST Women Fellowship.
\end{acknowledgments}

\bibliography{bibliography}

\end{document}